\documentclass[a4paper,12pt]{article}
\usepackage{amsmath}
\usepackage{amsfonts}
\usepackage{times}
\usepackage{amsthm}
\usepackage{epsfig}
\newcommand{\DATUM}{April 6, 2009}              
\newcommand{\comma}{\: ,}     
\newcommand{\period}{\: .}    
\newcommand{\Proof}{\noindent\emph{Proof. }}              
\newcommand{\QED}{\hspace*{\fill}\mbox{$\Box$}}           
\newcommand{\eps}{{\varepsilon}}
\newcommand{\e}{{\epsilon}}      
\newcommand{\vphi}{{\varphi}}           
\newcommand{\Om}{\Omega}                
\newcommand{\om}{\omega}
\newcommand{\la}{\langle}
\newcommand{\ra}{\rangle}
\newcommand{\C}{C_\Theta}
\newcommand{\one}{\mathbf{1}}

\newcommand{\cB}{\mathcal{B}}
\newcommand{\cD}{\mathcal{D}}

\newcommand{\cF}{\mathcal{F}}

\newcommand{\cH}{\mathcal{H}}

\newcommand{\cM}{\mathcal{M}}         
\newcommand{\cO}{\mathcal{O}}         

\newcommand{\cR}{\mathcal{R}}
\newcommand{\cS}{\mathcal{S}}
\newcommand{\cT}{\mathcal{T}}
\newcommand{\cU}{\mathcal{U}}

\newcommand{\cW}{\mathcal{W}}

\newcommand{\field}[1]{\mathbb{#1}}

\newcommand{\RR}{\field{R}}     
\newcommand{\NN}{\field{N}}     
\newcommand{\CC}{\field{C}}     
\newcommand{\fh}{\mathfrak{h}}  
\newcommand{\bchi}{{\overline{\chi}}}

\newcommand{\uw}{{\underline w}}
\newcommand{\umpnq}{{\underline{m,p,n,q}}}
\newcommand{\upq}{{\underline{0,p,0,q}}}
\newcommand{\tuw}{{\underline{\tilde{w}}}}
\newcommand{\huw}{{\underline{\hat{w}}}}

\newcommand{\hE}{\widehat{E}}
\newcommand{\hT}{\widehat{T}}    

\newcommand{\hw}{\hat{w}}

\newcommand{\tT}{\widetilde{T}} 
\newcommand{\tV}{\widetilde{V}} 
\newcommand{\tW}{\widetilde{W}} 
\newcommand{\tk}{\tilde{k}}

\newcommand{\tr}{\tilde{r}}
\newcommand{\tw}{\widetilde{w}}
\newcommand{\tx}{\tilde{x}}

\newcommand{\tpi}{\tilde{\pi}}

\newcommand{\sym}{\mathrm{sym}}
\newcommand{\const}{\mathrm{const}}
\newcommand{\red}{\mathrm{red}}
\newcommand{\rIm}{\mathrm{Im}}
\newcommand{\rRe}{\mathrm{Re}}               
\newcommand{\Ran}{\mathrm{Ran}}              
\newcommand{\spec}{\sigma}                   

\newcommand{\pointspec}{\sigma_{\rm pp}}

\newcommand{\cirS}{\mathop{\bigcirc\kern -.73em {\scriptstyle{\rm S}}}}

\newcommand{\dist}{{\rm dist}}

\newcommand{\dom}{\mathrm{dom}}
\newcommand{\supp}{\mathrm{supp}}

\newcommand{\cern}{\mathrm{Ker}}
\newcommand{\op}{\mathrm{op}}
\newcommand{\at}{{p}}

\newcommand{\hf}{H_f}

\newcommand{\adj}{\mathrm{ad}}
\newcommand{\comm}[2]{\left[#1,\,#2\right]}
\newcommand{\lb}{\left(}
\newcommand{\rb}{\right)}
\newcommand{\DETAILS}[1]{}
\renewcommand{\thesection}
{\Roman{section}}                      
\renewcommand{\theequation}
{\thesection.\arabic{equation}}        
\newcommand{\secct}[1]{\section{#1}
\setcounter{equation}{0}}              

\newtheorem{theorem}{Theorem}[section]         
\newtheorem{lemma}[theorem]{Lemma}             
\newtheorem{corollary}[theorem]{Corollary}     
\newtheorem{remark}[theorem]{Remark}           
\newtheorem{proposition}[theorem]{Proposition} 
\theoremstyle{plain}
\begin{document}
\bibliographystyle{plain}
\setcounter{page}{0}
\thispagestyle{empty}

\title{Spectral Renormalization Group and
Local Decay in the Standard Model of the Non-relativistic Quantum Electrodynamics}
\author{J\"{u}rg Fr\"{o}hlich
\thanks{Inst.~f.~Theoretische Physik; ETH Z\"{u}rich, Switzerland; also at IHES, Bures-sur-Yvette, France}
\and Marcel Griesemer
\thanks{Dept.~of Math.; Univ.~of Stuttgart; D-70569 Stuttgart, Germany}
%
\and Israel Michael Sigal
\thanks{School of Mathematics, IAS, Princeton, N.J., U.S.A.; permanent address: Dept.~of Math.; Univ. of Toronto; Toronto; Canada; Supported by NSERC Grant No. NA7901} \\
}
%

\date{\DATUM}

\maketitle

\begin{abstract}
  We prove the limiting absorption principle for the standard model of the
  non-relativistic quantum electrodynamics (QED) and for the Nelson model
  describing interactions of electrons with phonons. To this end we use
  the spectral renormalization group technique on the continuous spectrum in conjunction with
  the Mourre theory.

\end{abstract}

\thispagestyle{empty}
\setcounter{page}{1}

\secct{Introduction} \label{sec-I}

The mathematical framework of the theory of non-relativistic matter
interacting with the quantized electro-magnetic field
(non-relativistic quantum electrodynamics) is well established. It
is given in terms of the standard quantum Hamiltonian
\begin{equation}\label{eq1}
H^{SM}_g=\sum\limits_{j=1}^n{1\over 2m_j}
(i\nabla_{x_j}+gA(x_j))^2+V(x)+H_f
\end{equation}
acting on the Hilbert space $\cH=\cH_{p}\otimes\cH_{f}$, the tensor
product of the state spaces of the particle system and the quantized
electromagnetic field.  
Here $SM$ stands
for 'standard model'.
The notation above and units we use are explained below. This model
describes, in particular, the 
phenomena of emission and
absorption of radiation by systems of matter, such as atoms and
molecules, as well as other processes of interaction of quantum
radiation with matter. 
It has been extensively
studied in the last decade, see the books \cite{Spohn, GustafsonSigal} and reviews \cite{ Arai1999, Hirokawa1, Hirokawa2, Hiroshima5, Hiroshima7}  and references therein for a partial list of 
contributions.
%
%
%

For reasonable potentials $V(x)$ the operator $H^{SM}_g$ is
self-adjoint and its spectral and resonance structure - and
therefore dynamics for long but finite time-intervals - is well
understood (see e.g. \cite{AFFS,Faupin2007, GriesemerLiebLoss, HHS, HaslerHerbst2007, HaslerHerbstHuber2007, HirokawaHiroshimaSpohn, Sigal} and references therein for recent results). However, we still know little about its asymptotic
dynamics. In particular, the full scattering theory for this
operator does not, at present, exist (see, however,
\cite{FroehlichGriesemerSchlein1, FroehlichGriesemerSchlein2,
FroehlichGriesemerSchlein3, ChenFroehlichPizzo1, ChenFroehlichPizzo2}).

A key notion connected to the asymptotic dynamics is that of the
local decay. This notion also lies at the foundation of the
construction of the modern quantum scattering theory. It states that
the system under consideration is either in a bound state, or, as
time goes to infinity, it breaks apart, i.e. the
probability to occupy any bounded region of the physical space tends
to zero and, consequently, average distance between the particles
goes to infinity. 
In our case, this means that the
photons leave the part of the space occupied by the particle system.


Until recently the local decay for the Hamiltonian $H^{SM}_g$ is
proven only for the energies away from  $O(g^2)$-neighborhoods of
the ground state energy, $e_g$, and the ionization energy.
However, starting from any energy, the system eventually winds up in
a neighborhood of the ground state energy. Indeed, while the total
energy is conserved, the photons carry away the energy from regions
of the space where matter is concentrated. Hence understanding the
dynamics in this energy interval is an important matter.
Recently, the local decay was proven for states in
the spectral interval $(\e_g, \e_g + \e^{(p)}_{gap}/12)$ for the
Hamiltonian $H^{SM}_g$ \cite{FroehlichGriesemerSigal2008}.
%
Here $\e^{(p)}_{gap}:=\e^{(p)}_1 -\e^{(p)}_0$, where $\e^{(p)}_0$
and $\e^{(p)}_1$ are the ground state and the first excited state
energies of the particle system. In this paper we give another prove
of this fact.

However, the main goal of this paper is to develop a new approach to
time-dependent problems in the non-relativistic QED which
combines the spectral renormalization group (RG), developed in
\cite{BachFroehlichSigal1998a,BachFroehlichSigal1998b,BachChenFroehlichSigal2003} (see also \cite{FroehlichGriesemerSigal2009}),
with more traditional spectral techniques such as Mourre estimate. The key here is the result that the stronger property of the limiting absorption principle (LAP) propagates along the RG flow.


Now, we explain the units and notation employed in (\ref{eq1}). We use the units in
which the Planck constant divided by $2\pi$, speed of light and the
electron mass are equal to $1(\ \hbar=1$, $c=1$  and $m=1$). In
these units the electron charge is equal to $-\sqrt{\alpha}\
(e=-\sqrt{\alpha})$, where $\alpha =\frac{e^2}{4\pi \hbar c}\approx
{1\over 137}$ is the fine-structure constant, the distance, time and
energy are measured in the units of $\hbar/mc =3.86 \cdot
10^{-11}cm,\ \hbar/mc^2 =1.29 \cdot 10^{-21} sec$ and $mc^2 = 0.511
MeV$, respectively (natural units).
%
%
We show below that one can set $g:= \alpha^{3/2}$.

Our particle
system consists of $n$ particles of masses $m_j$ (the ratio of the
mass of the $j$-th particle to the mass of an electron) and
positions $x_j$, where $j=1, ..., n$. We write $x=(x_1,\dots,x_n)$.
The total potential of the particle system is denoted by $V(x)$. The
Hamiltonian operator of the particle system alone is given by
\begin{equation} \label{Hp}
H_p:=-\sum\limits_{j=1}^n {1\over 2m_j} \Delta_{x_j}+V(x),
\end{equation}
where $\Delta_{x_j}$ is the Laplacian in the variable $x_j$.  This
operator acts on a Hilbert space of the particle system, denoted by
$\cH_{p}$, which is either $L^2(\mathbb{R}^{3n})$ or a subspace of
this space determined by a symmetry group of the particle system.

The quantized electromagnetic field is described by the quantized (in
the Coulomb gauge) vector potential
\begin{equation}\label{eq3}
A(y)=\int(e^{iky}a(k)+e^{-iky}a^*(k)){\chi(k)d^3k\over (2\pi)^3
\sqrt{2|k|}},
\end{equation}
where $\chi$ is an ultraviolet cut-off: $\chi(k)=1$ in a
neighborhood of $k=0$ and it vanishes sufficiently fast at infinity,
and its dynamics, by the quantum Hamiltonian
\begin{equation} \label{Hf}
\hf \ = \ \int d^3 k \; a^*(k)\; \om(k) \; a(k) \comma
\end{equation}
both acting on the Fock space $\cH_{f}\equiv \cF$. Above,  $\om(k) \
= \ |k|$ is the dispersion law connecting the energy, $\omega(k)$,
of the field quantum with wave vector $k$,   $a^*(k)$ and $a(k)$
denote the creation and annihilation operators on $\cF$ and the
right side can be understood as a weak integral. The families
$a^*(k)$ and $a(k)$ are operator valued generalized, transverse
vector fields: $$a^\#(k):= \sum_{\lambda \in \{0, 1\}}
e_{\lambda}(k) a^\#_{\lambda}(k),$$ where $e_{\lambda}(k)$ are
polarization vectors, i.e. orthonormal vectors in $\mathbb{R}^3$
satisfying $k \cdot e_{\lambda}(k) =0$, and $a^\#_{\lambda}(k)$ are
scalar creation and annihilation operators satisfying standard
commutation relations. See Supplement for a brief review of the
definitions of the Fock space, the creation and annihilation
operators acting on it and the definition of the operator $\hf$.

In the natural units the Hamiltonian operator
is of the form \eqref{eq1} but with $g=e$ and with $V(x)$ being the
total Coulomb potential of the particle system.
To obtain expression (\ref{eq1}) we rescale this original
Hamiltonian appropriately (see \cite{BachFroehlichSigal1999}). Then
we relax the restriction on $V(x)$ and consider standard generalized
$n$-body potentials (see e.g. \cite{HunzikerSigal}), $V(x) = \sum_i
W_i(\pi_i x)$, where $\pi_i$ are a linear maps from
$\mathbb{R}^{3n}$ to $\mathbb{R}^{m_i},\ m_i \le 3n $ and $ W_i$ are
Kato-Rellich potentials (i.e. $W_i(\pi_i x)  \in
L^{p_i}(\mathbb{R}^{m_i}) + (L^\infty (\mathbb{R}^{3n}))_\eps$ with
$p_i=2$ for $m_i \le 3,\ p_i>2 $ for $m_i =4$ and $p_i \ge m_i/2$
for $m_i > 4$, see \cite{RSIV, HislopSigal}). In order not to deal
with the problem of center-of-mass motion, which is not essential in
the present context, we assume that either some of the particles
(nuclei) are infinitely heavy or the system is placed in an external
potential field.

One verifies that $\hf$ defines a positive, self-adjoint operator on
$\cF$ with purely absolutely continuous spectrum, except for a
simple eigenvalue $0$ corresponding to the eigenvector $\Om$ (the
vacuum vector, see Supplement). Thus for $g=0$ the low energy
spectrum of the operator $H^{SM}_0$
consists of the branches $[\e^{(p)}_i, \infty)$, where $\e^{(p)}_i$
are the isolated eigenvalues of $H_\at$, and of the eigenvalues
$\e^{(p)}_i$ sitting at the top of the branch points ('thresholds')
of the continuous spectrum. The absence of gaps between the
eigenvalues and thresholds is a consequence of the fact that the
photons (and the phonons) are massless. This leads to hard and
subtle problems in the perturbation theory, known collectively as
the infrared problem.

As was mentioned above, in this paper we prove the local decay
property for the Hamiltonian $H^{SM}_g$.
In fact, we prove a slightly stronger property - the limiting
absorption principle - which states that the resolvent sandwiched by
appropriate weights has H\"older continuous limit on the spectrum.
To be specific, let $B$ denote the self-adjoint generator of
dilatations on the Fock space $\cF$. It can be expressed in terms of
creation- and annihilation operators as
\begin{equation} \label{eq-I.24}
B \ = \ \frac{i}{2}\; \int d^3k \; a^*(k) \: \big\{ k \cdot
\nabla_k + \nabla_k \cdot k \big\} \: a(k) \period
\end{equation}
We further extend it to the Hilbert space $\cH=\cH_{\at}\otimes\cF$.
Let $\la B\ra := (\one +B^2)^{1/2}$.
Our goal is to prove the following\\
%
\begin{theorem} \label{main-thm} Let
$g\ll \e^{(p)}_{gap}$ and let $\Delta \subset (\e_g, \e_g
+\frac{1}{12}\ \e^{(p)}_{gap}) $, where $\e_g$ is the ground state
energy of $H$. Then
\begin{equation}
\la B\ra^{-\theta}(H^{SM}_g-\lambda \pm i 0)^{-1}\la B \ra^{-\theta}\in
C^\nu(\Delta). \label{eqn:Claim1}
\end{equation}
for $\theta>1/2$ and $0<\nu<\theta-\frac{1}{2}$.
\end{theorem}

The above theorem has the following consequence. (In what follows
functions of self-adjoint operators are defined by functional calculus.)\\
\begin{corollary}
For $\Delta$ as above and for any function $f(\lambda)$ with $\supp
f \subseteq\Delta$ and for $\nu< \theta-\frac{1}{2}$, we have
\begin{equation}
\|\la B\ra^{-\theta}e^{-i H t}f(H)\la B\ra^{-\theta}\|\le C
t^{-\nu}.
\end{equation}
\end{corollary}
The statement follows from \eqref{eqn:Claim1} and the formula
\begin{align*}
& \la B\ra^{-\theta}e^{-i H t}f(H)\la B\ra^{-\theta}= &
\int_{-\infty}^\infty d\lambda f(\lambda)e^{-i\lambda t}\rIm\la
B\ra^{-\theta}(H-\lambda-i0)^{-1}\la B\ra^{-\theta}
\end{align*}
(see e.g. \cite{RSIV} and a detailed discussion in
\cite{FroehlichGriesemerSigal2008}).

\begin{remark} \label{rem-1} Let $\Sigma_{p}:=\inf\sigma(H_{p})$. We expect that
the method of this paper can be extended to the energy
interval $\spec(H)\setminus \pointspec(H)$ for the Nelson model and
$\big(\spec(H)\setminus \pointspec(H)\big) \bigcap
(-\infty,\Sigma_{p}-\eps]$ for some $\eps>0$, for QED.
\end{remark}

Previously the limiting absorption principle and local decay
estimates were proven in \cite {BachFroehlichSigal1998b, BachFroehlichSigalSoffer1999} for
the standard model of non-relativistic QED and for the Nelson model
away from neighborhoods of the ground state energy and ionization
threshold. In \cite {GergescuGerardMoeller1,GergescuGerardMoeller2}
they were proven  for the Nelson model near the ground state energy
and for all values of the coupling constant, but under rather
stringent assumptions, including that on the infra-red behavior of
the coupling functions (see also
\cite{BachFroehlichSigal1998a,BachFroehlichSigal1999,HuebnerSpohn2,Skibsted}
for earlier works). Finally, as was mentioned above, it was proven in a neighbourhood of the ground state energy in \cite{FroehlichGriesemerSigal2008}.

Our approach consists of three steps. First, following \cite{Sigal},
we use a generalized Pauli-Fierz transform to map the QED
Hamiltonian  Eq.~(\ref{eq1}) into a new Hamiltonian $H^{PF}_g$ whose
interaction has a better, in some sense, infra-red behaviour. To
this new Hamiltonian we apply sufficiently many iterations of the
renormalization map obtaining at the end a rather simple Hamiltonian
which we investigate further with the help of the Mourre estimate.
This proves LAP for the latter Hamiltonian. Since, as we prove in
this paper, the renormalization map preserves the LAP property we
conclude from this that the Hamiltonian $H^{PF}_g$ enjoys the LAP
property as well. The size of the interval and the number of iterations of the RG map depends on the distance of a spectral point of interest to the ground state energy.




\DETAILS{Note that in this approach a specific form of the interaction and
the coupling functions becomes irrelevant.
With little more work one can establish an explicit restriction on
the coupling constant $g$ in terms of the particle energy difference
$\e^{(p)}_{gap}$ and appropriate norms of the coupling functions.}

In this paper we consider also the Nelson model. In
that model the total system consisting of the particle system
coupled to the quantized field is described by the Hamiltonian
\begin{equation} \label{Hn}
H_g^{N} \ = \ H^{N}_0 \, + \, I_{g}^{N} \comma
\end{equation}
acting on the state space, $\cH=\cH_\at \, \otimes \, \cF$, where
now $\cF$ is the  Fock space for phonons, i.~e. spinless, massless
Bosons.
%
%
Here $g$ is a positive parameter - a coupling constant - which we
assume to be small, and
\begin{equation} \label{H_0}
H^{N}_0 \ = \ H^{N}_p \, + \, \hf \comma
\end{equation}
where $H^{N}_p=H_p$ and $\hf$ are given in \eqref{Hp} and
\eqref{Hf}, respectively,  but, in the last case, with the scalar
creation and annihilation operators, $a$ and $a^*$, and the
interaction operator is
%
\begin{equation} \label{I.6}
I_g \ = \ g \, \int \frac{ d^3 k \: \kappa(k)}{ |k|^{1/2} } \:
\big\{ e^{-i k x} \, a^*(k) \: + \: e^{i k x} \, a(k) \big\}
\end{equation}
(we can also treat terms quadratic in $a$ and $a^*$ but for the sake
of exposition we leave such terms out). Here, $\kappa=\kappa(k)$ is
a real function with the property that
\begin{equation} \label{I.7}
\| \kappa \|_\mu \ := \ \Big( \int \frac{d^3 k}{|k|^{3+2\mu} } \:
|\kappa(k)|^2 \Big)^{1/2} \ < \ \infty \comma
\end{equation}
for some (arbitrarily small, but) strictly positive $\mu > 0$.
In the following, we fix $\kappa$ with $\|\kappa\|_{\mu} = 1$ and
vary $g$. It is easy to see that the operator $I_g$ is symmetric and
bounded relative to $H^{N}_0$, in the sense of Kato \cite{RSIV, HislopSigal},
with an arbitrarily small constant. Thus $H_g^{N}$ is self-adjoint
on the domain of $H^{N}_0$ for arbitrary $g$.

Of course, for the Nelson model we can take an arbitrary dimension
$d\geq 1$ rather than the dimension $3$.
Our approach can handle the interactions quadratic in creation and
annihilation operators, $a$ and $a^*$, as it is the case for the
operator $H^{SM}_g$. All the results mentioned above for the
standard model Hamiltonian $H^{SM}_g$ holds also for the Nelson
model one, $H^{N}_g$ with $\mu >0$.


In order not complicate matters unnecessary we will think about the
creation and annihilation operators used below as scalar operators
rather than operator-valued transverse vector functions. We explain
at the end of Appendix how to reinterpret the corresponding
expression for the vector - photon - case.

\secct{Generalized Pauli-Fierz Transform} \label{sec-II}
%
We describe the generalized Pauli-Fierz transform mentioned in the
introduction (see \cite{Sigal}). We define the following Hamiltonian
\begin{equation}
H_g^{PF}: = e^{-ig F(x)} H^{SM}_g e^{ig F(x)},
\end{equation}

\noindent which we call the generalized Pauli-Fierz Hamiltonian. In
order to keep notation simple, we present this transformation in the
one-particle, $n=1$, case:
%
\begin{equation}\label{eq3}
F(x)=\sum_\lambda
\int(\bar{f}_{x,\lambda}(k)a_{\lambda}(k)+f_{x,\lambda}(k)a_{\lambda}^*(k))\frac{d^3k}{\sqrt{|k|}},
\end{equation}
with the coupling function $f_{x,\lambda}(k)$ chosen as
$$f_{x,\lambda}(k):=
\frac{e^{-ikx}\chi(k)}{(2\pi)^3\sqrt{2|k|}}\varphi(|k|^{\frac{1}{2}}\eps_\lambda(k)
\cdot x).$$ The function $\varphi$ is assumed to be $C^2$, bounded, having a bounded second derivative 
and satisfying $\varphi'(0)=1.$ We compute


\begin{equation} \label{H^PFa}
H_g^{PF} = \frac{1}{2} (p - g A_1(x))^2 + V_g(x) + H_f + gG(x)\\
\end{equation}

\noindent where $A_1(x) = A(x) - \nabla F(x),\ V_g(x):= V(x) + 2
g^2\sum_\lambda \int \omega |f_{x,\lambda}(k)|^2d^3k$ and
\begin{equation} \label{I.14}
G(x):=- i\sum_\lambda
\int\omega(\bar{f}_{x,\lambda}(k)a_{\lambda}(k)-f_{x,\lambda}(k)a_{\lambda}^*(k))\frac{d^3k}{\sqrt{|k|}}.
\end{equation}
%
%
(The terms $gG$ and $V_g - V$ come from the commutator expansion
$e^{-ig F(x)} H_f e^{ig F(x)}$ $= - i g [F,H_f] - g^2 [F, [F,
H_f]]$.) Observe that the operator family $A_1(x)$ is of the form
\begin{equation}\label{15}
A_1(x)=\sum_\lambda
\int(e^{ikx}a_{\lambda}(k)+e^{-ikx}a_{\lambda}^*(k)){\chi_{\lambda,
x}(k)d^3k\over (2\pi)^3 \sqrt{2|k|}},
\end{equation}
where the coupling function 
$\chi_{\lambda, x}(k)$ is defined as follows $$\chi_{\lambda, x}(k):=
e_{\lambda}(k)e^{-ikx}\chi(k)- \nabla_x f_{x,\lambda}(k).$$
It satisfies the estimates
\begin{equation}\label{Acoupling}
|\chi_{\lambda, x}(k)|
\le \const \min (1, \sqrt{|k|}\la x\ra),
\end{equation}
with $\la x\ra := (1 +|x|^2)^{1/2}$, and
\begin{equation}\label{chi-estim2}
\int \frac{d^3 k}{|k| } \: |\chi_{\lambda, x}(k)|^{2}  \ < \ \infty
.\end{equation}

Using the fact that the operators $A_1$ and $G$ have much better
infra-red behavior than the original vector potential $A$ we can use
our approach and prove the limiting absorption principle for
$H_g^{PF}$ and $B$:
\begin{equation} \label{lap1}
\langle B \rangle^{-\theta} (H_g^{PF} - z)^{-1}\langle B
\rangle^{-\theta} \text{ H\"older continuous in $z$}.
\end{equation}

Now we show that estimate \eqref{lap1} and the additional
restriction on the spectral interval imply the limiting absorption
principle for $H^{SM}_g$.  Let $B_1 : = e^{-ig F(x)} B e^{ig F(x)}$.
We compute


\begin{equation}
B_1
= B + g C
\end{equation}

\noindent where $C : = -i[F(x), B]$.  Note that the operator $C$
contains a term proportional to $x$. Now,
let a function $f$ be supported in $(-\infty,\Sigma_{p})$. Then,
using that $(H^{SM}_g - z)^{-1}= e^{ig F(x)}(H_g^{PF} - z)^{-1}
e^{-ig F(x)}$, we obtain


\begin{align}
&\langle B\rangle^{-\theta}f(H^{SM}_g)^2 (H^{SM}_g - z)^{-1}\langle B\rangle^{-\theta}\\
\notag
&= D E(z) D^*,
\end{align}
where $ D := \langle B\rangle^{-\theta}f(H^{SM}_g)\langle B_1
\rangle^{\theta} e^{ig F(x)}$
and $ E(z):= \langle B\rangle^{-\theta}(H_g^{PF} - z)^{-1}\langle
B\rangle^{-\theta}$. The operator $D$ is bounded by standard
operator calculus estimates and the fact that $e^{\delta \langle x
\rangle} f(H^{SM}_g)$ is bounded for $\delta >0$ sufficiently small.
Furthermore, the operator-family $E(z)$ is H\"older continuous by
the assumed result.
Now, for $z\in (-\infty,\Sigma_{p}-\eps]$ for some $\eps>0$ the
previous conclusion remains true even if remove the cut-off function
$f(H^{SM}_g)$.

We mention for further references that the operator (I.13) can be
written as
\begin{equation} \label{Hpf}
H_g^{PF} \ = \ H^{PF}_{0} \, + \, I_{g}^{PF} \comma
\end{equation}
where $H_{0g}=H_0 + 2 g^2\sum_\lambda \int \omega
|f_{x,\lambda}(k)|^2d^3k + g^2\sum_\lambda
\int{|\chi_\lambda(k)|^2\over (2\pi)^6 2\omega}d^3k$, with $H_0$
defined in (\ref{H_0}), and $I_{g}^{PF}$ is defined by this
relation. Note that the operator $I_{g}^{PF}$ contains linear and
quadratic terms in the creation and annihilation operators, with the
coupling functions (form-factors) in the linear terms satisfying
estimate \eqref{Acoupling}  and with the coupling functions in the
quadratic terms satisfying a similar estimate. Moreover, the operator
$H^{PF}_{0}$ is of the form $H^{PF}_{0}= H^{PF}_{p}+H_{f}$ where
\begin{equation} \label{Hpfp}
H^{PF}_{p}:=H_{p}+2 g^2\sum_\lambda \int |k|
|f_{x,\lambda}(k)|^2d^3k + g^2\sum_\lambda
\int{|\chi_\lambda(k)|^2\over |k|}d^3k
\end{equation}
where $H_p$ is given in \eqref{Hp}.

\secct{The Smooth Feshbach-Shur Map} \label{sec-III}
%
In this section, we review and extend ,in a simple but important
way, the method of isospectral decimations or Feshbach-Schur maps
introduced in \cite{BachFroehlichSigal1998a,BachFroehlichSigal1998b}
and refined in \cite{BachChenFroehlichSigal2003}\footnote{In
\cite{BachFroehlichSigal1998a,BachFroehlichSigal1998b,
BachChenFroehlichSigal2003} this map is called the Feshbach map. As
was pointed out to us by F. Klopp and B. Simon, the invertibility
procedure at the heart of this map was introduced by I. Schur in
1917; it appeared implicitly in an independent work of H. Feshbach
on the theory of nuclear reactions in 1958, where the problem of
perturbation of operator eigenvalues was considered. See
\cite{GriesemerHasler} for further discussion and historical
remarks.}. For further extensions see  \cite{GriesemerHasler}.
At the root of this method is the isospectral smooth Feshbach-Schur
map acting on a set of closed operators and mapping a given operator
to one  acting on much smaller space which is easier to handle.


%
%
Let $\chi$,  $\bchi$ be a partition of unity on a separable Hilbert
space $\cH$, i.e. $\chi$ and  $\bchi$ are positive operators on
$\cH$ whose norms are bounded by one, $0 \leq \chi, \bchi \leq
\mathbf{1}$, and $\chi^{2}+ \bchi^{2} = \mathbf{1}$. We assume that
$\chi$ and $\bchi$ are nonzero. Let $\tau$ be a (linear) projection
acting on closed operators on $\cH$ s.t. operators from its image
commute with $\chi$ and $\bchi$. We also assume that
$\tau(\textbf{1}) =\textbf{1}$.
Assume that $\tau$ and $\chi$ (and therefore also $\bchi$) leave
$\dom(H)$ invariant $ \dom(\tau(H))=\dom(H)$ and $
\chi\dom(H)\subset \dom(H) $. Let $\overline{\tau}:= \mathbf{1} -
\tau$ and define
\begin{equation}
\\ \label{II-1}
H_{\tau,\chi^{\#}} \ \; :=  \tau(H) \: + \: \chi^{\#}
\overline{\tau}(H)\chi^{\#} \period
\end{equation}
where $\chi^{\#}$ stands for either $\chi$ or $\bchi$.

Given $\chi$ and $\tau$ as above, we denote by $D_{\tau,\chi}$ the
space of closed operators, $H$, on $\cH$ which belong to the domain
of $\tau$ and satisfy the following conditions:
\begin{equation}
\label{II-2}H_{\tau,\bchi}\ \mbox{is (bounded) invertible on}\ \Ran
\, \bchi,
\end{equation}
$$\overline{\tau}(H) \chi\ \mbox{and}\ \chi \overline{\tau}(H)\
\mbox{extend to bounded operators on}\ \cH.$$ (For more general
conditions see \cite{BachChenFroehlichSigal2003, GriesemerHasler}.)
%
%

Denote $H_0 := \tau(H)$ and $W := \overline{\tau}(H)$. Then $H_0$
and $W$ are two closed operators on $\cH$ with coinciding domains, $
\dom(H_0)= \dom(W)=\dom(H)$, and $H = H_0 + W$. We remark that
the domains of $\chi W\chi$, $\bchi W\bchi$, $H_{\tau,\chi}$,
and $H_{\tau,\bchi}$ all contain $\dom(H)$.

The \textit{smooth Feshbach-Schur map (SFM)} maps operators on $\cH$
to operators on $\cH$ by $H \ \mapsto \ F_{\tau,\chi} (H)$, where
\begin{equation} \label{II-3}
 F_{\tau,\chi} (H) \ := \ H_0 \, + \, \chi W\chi \, -
\, \chi W \bchi H_{\tau,\bchi}^{-1} \bchi W \chi \period
\end{equation}
Clearly, it is defined on the domain $D_{\tau,\chi}$.

Remarks

\begin{itemize}
\item The definition of the smooth Feshbach-Schur map
given above is the same as in \cite{FroehlichGriesemerSigal2009} and differs from the one given in
\cite{BachChenFroehlichSigal2003}. In
\cite{BachChenFroehlichSigal2003} the map $F_{\tau,\chi} (H)$ is
denoted by $F_{\chi}(H,\tau(H))$ and the pair of operators $(H, T)$
are referred to as a Feshbach pair.

\item The Feshbach-Schur map is obtained from the smooth Feshbach-Schur map by
specifying  $\chi=$ projection and, usually, $\tau = 0$.


\end{itemize}
We furthermore define the maps entering some identities involving
the Feshbach-Schur map:
\begin{eqnarray} \label{eq-II-4}
Q_{\tau,\chi} (H) & := & \chi \: - \: \bchi \, H_{\tau,\bchi}^{-1}
\bchi W \chi \comma
\\  \label{eq-II-5}
Q_{\tau,\chi} ^\#(H) & := & \chi \: - \: \chi W \bchi \,
H_{\tau,\bchi}^{-1} \bchi \period
\end{eqnarray}
Note that $Q_{\tau,\chi} (H) \in \cB( \Ran\, \chi , \cH)$ and
$Q_{\tau,\chi}^\#(H) \in \cB( \cH , \Ran\, \chi)$.

The smooth Feshbach map of $H$ is isospectral to $H$ in the sense
of the following theorem.
%
\begin{theorem} \label{thm-II-1}
Let $\chi$ and $\tau$ be as above. Then we have the following
results.
\begin{itemize}
\item[(i)] $0 \in \rho(H) \leftrightarrow 0 \in \rho(F_{\tau,\chi} (H))$, i.e. $H$ is bounded invertible on $\cH$ if and only if
$F_{\tau,\chi} (H)$ is bounded invertible on $\Ran\, \chi$.
\item[(ii)] If $\psi \in \cH \setminus \{0\}$ solves $H \psi = 0$
then $\vphi := \chi \psi \in \Ran\, \chi \setminus \{0\}$ solves
$F_{\tau,\chi} (H) \, \vphi = 0$. \item[(iii)] If $\vphi \in
\Ran\, \chi \setminus \{0\}$ solves $F_{\tau,\chi} (H) \, \vphi =
0$ then $\psi := Q_{\tau,\chi} (H) \vphi \in \cH \setminus \{0\}$
solves $H \psi = 0$. \item[(iv)] The multiplicity of the spectral
value $\{0\}$ is conserved in the sense that $\dim \cern H = \dim
\cern F_{\tau,\chi} (H)$.
\item[(v)] If one of the inverses, $H^{-1}$ or $F_{\tau,\chi} (H)^{-1}$, exists then so is the other
and they are related as
\begin{equation} \label{eq-II-6}
H^{-1}  =  Q_{\tau,\chi} (H) \: F_{\tau,\chi} (H)^{-1} \:
Q_{\tau,\chi} (H)^\# \; + \; \bchi \, H_{\tau,\bchi}^{-1} \bchi
\comma
\end{equation}
and $$ F_{\tau,\chi} (H)^{-1}  =  \chi \, H^{-1} \, \chi \; + \;
\bchi \, T^{-1} \bchi \period$$
%
\end{itemize}
\end{theorem}

This theorem is proven in \cite{BachChenFroehlichSigal2003} (see
\cite{GriesemerHasler} for a more general result). Now we establish
a key result relating smoothness of the resolvent of an operator on
its continuous spectrum with smoothness of the resolvent of its
image under a smooth Feshbach-Schur map. Let $B_\theta:=\la B\ra^{-\theta}$.
In what follows $\Delta$ stands for an open interval in $\mathbb{R}$.
\begin{theorem} \label{LAPtransfer}
Assume a self-adjoint operator $B$ and a $C^\infty$ family
$H(\lambda)$, $\lambda\in\Delta$, of closed operators satisfy the
following conditions: $\forall \lambda\in\Delta, H_{\tau,\bchi}(\lambda) \in
D_{\tau,\chi}$ and
\begin{equation}
\adj_B^j(A)\ \mbox{is bounded}\ \forall j\le 1, \label{eqn:5}
\end{equation}
where $A$ stands for one of the operators
$A=\chi,\ \overline{\chi},\ \chi W,\ W\chi,\ \partial_\lambda^k (\bchi
H_{\tau,\bchi}(\lambda)^{-1} \bchi)\
\forall k.$
If $H(\lambda)\in \dom(F_{\tau, \chi})$, then for any $\nu\ge 0$ and $0 < \theta \le 1$,
\begin{equation}
B_\theta(F_{\tau, \chi}(H(\lambda))-i0)^{-1}B_\theta\in
C^\nu(\Delta)\Rightarrow B_\theta(H(\lambda)-i0)^{-1}B_\theta\in
C^\nu(\Delta). \label{eqn:7}
\end{equation}
\end{theorem}
\Proof  We use identity (\ref{eq-II-6}) with $H$ replaced by
$H(\lambda) -i\eps$. Since $\tau(\textbf{1})=\textbf{1}$ we have
that
$(H(\lambda)-i\epsilon)_{\tau,\chi^{\#}}=H(\lambda)_{\tau,\chi^{\#}}-i\epsilon$,
where $\chi^{\#}$ is either $\chi$ or $\bchi$. Furthermore, on
$Ran \bchi$, the operator family
$[(H(\lambda)-i\eps)_{\tau,\bchi}]^{-1}$ is differentiable in $\lambda$ and analytic in $\eps$ and
can be expanded as
$$[(H(\lambda)-i\eps)_{\tau,\bchi}]^{-1}=
[H(\lambda)_{\tau,\bchi}]^{-1} + i\eps
[H(\lambda)_{\tau,\bchi}]^{-1} \bchi^2
[H(\lambda)_{\tau,\bchi}]^{-1} +O(\eps^2).$$ This implies the
relation \begin{equation} \label{BFB}
\lim_{\eps\rightarrow
i0}B_\theta[F_{\tau,\chi}(H(\lambda)-i\eps)]^{-1}B_\theta =
B_\theta[F_{\tau,\chi}(H(\lambda))-i0]^{-1}B_\theta.
\end{equation}
Conditions \eqref{eqn:5} and
the formula $B_\theta=C_\theta\int_0^\infty
\frac{d\omega}{\omega^{\theta/2}}(\omega+1+B^2)^{-1}$, where
$C_\theta:=\big[\int_0^\infty
\frac{d\omega}{\omega^{\theta/2}}(\omega+1)^{-1}\big]^{-1}$, imply
that the operators
\begin{equation}
B_\theta\chi B_\theta^{-1}, B_\theta\overline{\chi} B_\theta^{-1},
B_\theta [H(\lambda)_{\tau,\bchi}]^{-1}B_{\theta}^{-1}
\end{equation}
and the transposed operators (i.e., $B_\theta^{-1}\chi B_\theta$,
etc.) are bounded and $C^\infty(\Delta)$ in $\lambda$.
This property shows that  $B_\theta^{-1}Q B_\theta$ and $B_\theta^{-1} Q^\# B_\theta$  are bounded and smooth in $\lambda \in \Delta$. This together with \eqref{BFB}, $H(\lambda)\in \dom(F_{\tau\chi})$ and \eqref{eq-II-6} 
implies the theorem.\qed
\secct{A Banach Space of Hamiltonians} \label{subsec-III.1}
%
We construct a Banach space of Hamiltonians on which the
renormalization transformation is defined.
%
%
Let $\chi_1(r) \equiv\chi_{r\le1}$ be a smooth cut-off function s.t.
$\chi_1 = 1$ for $r \le 1,\ = 0$ for $r\ge 11/10$ and $0 \le
\chi_1(r) \le1\ $ and $\sup|\partial^n_r \chi_1(r)| \le 30\ \forall
r$ and for $n=1,2.$ We denote $\chi_\rho(r) \equiv\chi_{r\le\rho}:=
\chi_1(r/\rho) \equiv\chi_{r/\rho\le1}$ and
$\chi_\rho\equiv\chi_{H_f\le\rho}$.

Let $B_1^d$ denotes the unit ball in $\RR^{3d}$, $I:=[0,1]$ and $m,n
\ge 0$. Given functions $w_{0,0}: [0, \infty) \rightarrow
\mathbb{C}$ and $w_{m,n}: I\times B_1^{m+n} \rightarrow \mathbb{C},
m+n > 0$, we consider monomials, $W_{m,n}  \equiv W_{m,n}[w_{m,n}]$,
in the creation and annihilation operators of the form
%
%
%
$W_{0,0}:=w_{0,0}[H_f]$ (defined by the operator calculus),  for
$m=n=0$, and
\begin{eqnarray} \label{III.1}
&&W_{m,n}[w_{m,n}]  :=
\\ \nonumber
&&
\int_{B_1^{m+n}} \frac{ dk_{(m,n)} }{ |k_{(m,n)}|^{1/2} } \; a^*(
k_{(m)} ) \, w_{m,n} \big[ \hf ; k_{(m,n)} \big] \, a( \tk_{(n)} )
\:,
\end{eqnarray}
%
for $m+n>0$. Here we used the notation
\begin{eqnarray} \label{III.2}
& k_{(m)} \: := \: (k_1, \ldots, k_m) \: \in \: \RR^{dm} \comma
\hspace{5mm}
a^*( k_{(m)} ) \: := \: \prod_{i=1}^m a^*(k_i ),
\\  \label{III.3}
& k_{(m,n)} \: := \: (k_{(m)}, \tk_{(n)}) \comma \hspace{5mm}
dk_{(m,n)} \: := \: \prod_{i=1}^m  d^d k_i \; \prod_{i=1}^n d^d
\tk_i \comma &
\\  \label{III.4}
& |k_{(m,n)}| \, := \, |k_{(m)}| \cdot |\tk_{(n)}| \comma
\hspace{3mm} |k_{(m)}| \, := \, |k_1| \cdots |k_m| \period &
\end{eqnarray}

We assume that for every $m$ and $n$ with $m+n>0$ the function
$w_{m,n}[ r, , k_{(m,n)}]$
is $s$ times continuously differentiable in $r \in I$, for almost
every $k_{(m,n)} \in B_1^{m+n}$, and weakly differentiable in
$k_{(m,n)} \in B_1^{m+n}$, for almost every $r$ in $I$. As a
function of $k_{(m,n)}$, it is totally symmetric w.~r.~t.\ the
variables $k_{(m)} = (k_1, \ldots, k_m)$ and $\tk_{(n)} = (\tk_1,
\ldots, \tk_n)$ and obeys the norm bound
\begin{equation} \label{III.5}
\| w_{m,n} \|_{\mu,s} \ :=
\sum \|  \partial_r^n (k\partial_k)^q w_{m,n} \|_{\mu} \ < \ \infty
\comma
\end{equation}
where $q:= (q_1,  \ldots, q_{m+n}),\ (k\partial_k)^q: =
\prod_1^{m+n}(k_j \cdot \nabla_{k_j})^{q_j}$, with $k_{m+j} :=
\tk_j$, and where  the sum is taken over the indices $n$ and $q$
satisfying $0 \le n+|q| \leq s$ and where $\mu \ge 0$ and
%
%
%
\begin{equation} \label{III.6}
\| w_{m,n} \|_{\mu} \ := \max_j \sup_{r \in I, k_{(m,n)} \in
B_1^{m+n}} \big| | k_j|^{-\mu}w_{m,n}[r ; k_{(m,n)}] \big|.
\end{equation}
Here and in what follows $k_j$ is the $j-$th $3-$dimensional
components of the $k-$vector $k_{(m,n)}$ over we take the supremum.
For $m+n=0$ the variable $r$ ranges in $[0,\infty)$ and we assume
that the following norm is finite:
\begin{equation}
\\ \label{III.7}
\ \| w_{0,0} \|_{\mu, s} := |w_{0,0}(0)|+ \sum_{1 \le n \leq s}
\sup_{r \in I}|
\partial_r^n w_{0,0}(r)|
 \hspace{10mm}
\end{equation}
(for $s=0$ we drop the sum on the r.h.s. ). (This norm is
independent of $\mu$ but we keep this index for notational
convinience.) The Banach space of these functions is denoted by
$\cW_{m,n}^{\mu,s}$. Moreover, $W_{m,n}[w_{m,n}]$ stresses the
dependence of $W_{m,n}$ on $w_{m,n}$.
%
%
%
%
In particular, $W_{0,0}[w_{0,0}] := w_{0,0}[\hf]$.

We fix three numbers $\mu$, $0 < \xi < 1$ and $s \ge 0$ and define
Banach space
\begin{equation} \label{III.8}
\cW^{\mu,s} \ \equiv \cW^{\mu,s}_{\xi} := \ \bigoplus_{m+n \geq 0}
\cW_{m,n}^{\mu,s} \ \comma
\end{equation}
with the norm
\begin{equation} \label{III.9}
\big\|  \uw \big\|_{\mu, s,\xi} \ := \ \sum_{m+n \geq 0}
\xi^{-(m+n)} \; \| w_{m,n} \|_{\mu, s} \ < \ \infty \period
\end{equation}
Clearly, $\cW^{\mu,s'}_{\xi', \mu'} \subset \cW^{\mu,s}_{\xi,\mu}$
if $\mu' \le \mu, s' \ge s$ and $\xi' \le \xi$.

\begin{remark} \label{remIII-1} Though we use the same notation, the
Banach spaces, $\cW^{\mu,s}_{\xi,\mu}$, etc, introduced above differ
from the ones used in \cite{Sigal, FroehlichGriesemerSigal2007b}.
The latter are obtained from the former by setting $q=0$ in
\eqref{III.5}. To extend estimates of \cite{Sigal,
FroehlichGriesemerSigal2007b} to the present setting one has to
estimate the effect of the derivatives $(k\partial_k)^q$ which is
straightforward.
\end{remark}

The following basic bound, proven in [2], links the norm defined in
(\ref{III.6})
to the operator norm on
$\cB[\cF]$.
%
\begin{theorem} \label{thm-III.1}
Fix $m,n \in \NN_0$ such that $m+n \geq 1$. Suppose that $w_{m,n}
\in \cW_{m,n}^{\mu,0}$, and let $W_{m,n} \equiv W_{m,n}[w_{m,n}]$ be
as defined in (\ref{III.1}). Then $\forall \rho>0$
%
\begin{equation} \label{III.10}
\big\|  (\hf+\rho)^{-m/2} \, W_{m,n} \,
(\hf+\rho)^{-n/2}
\big\|
\ \leq \  \| w_{m,n} \|_{0}
\, ,
\end{equation}
and therefore
\begin{equation} \label{III.11}
\big\| \chi_\rho \, W_{m,n} \,
\chi_\rho
 \big\|
\ \leq \ \frac{\rho^{(m+n)(1+\mu)}}{\sqrt{m! \, n!} } \, \| w_{m,n}
\|_{\mu} \, ,
\end{equation}
where $\| \, \cdot \, \|$ denotes the operator norm on $\cB[\cF]$.
\end{theorem}

Theorem~\ref{thm-III.1} says that the finiteness of $\| w_{m,n}
\|_{\mu}$ insures that $W_{m,n}$ defines a bounded operator on
$\cB[\cF]$.

Now with a sequence $\uw := (w_{m,n})_{m+n \geq 0}$ in $\cW^{\mu,s}$
we associate an operator
by setting
\begin{equation} \label{III.12}
H(\uw)  := W_{0,0}[\uw] + \sum_{m+n \geq 1} \chi_1 W_{m,n}[\uw]
\chi_1,
\end{equation}
where we write $W_{m,n}[\uw] := W_{m,n}[w_{m,n}]$. This form of
operators on the Fock space will be called the \textit{generalized
normal (or Wick) form}. Theorem~\ref{thm-III.1} shows that the
series in (\ref{III.12}) converges in the operator norm and obeys
the
estimate
\begin{equation} \label{eq-III-1-25.1}
\big\| \, H(\uw)- W_{0,0}(\uw) \, \big\| \ \leq \ \xi\big\| \, \uw_1
\, \big\|_{\mu,0, \xi} \comma
\end{equation}
for any $\uw = (w_{m,n})_{m+n \geq 0} \in \cW^{\mu,0}$.  Here $\uw_1
= (w_{m,n})_{m+n \geq 1}$. Hence we have the linear map
\begin{equation} \label{eq-III-1-24.1}
H : \uw \to H(\uw)
\end{equation}
from $\cW^{\mu,0}$ into the set of closed operators on the Fock
space $\cF$.
Furthermore the following result was proven in [2].
%
\begin{theorem} \label{thm-III-1-2}
For any $\mu \ge 0$ and $0 < \xi < 1$, the map $H : \uw \to H(\uw)$,
given in (\ref{III.12}), is one-to-one.
%
%
\end{theorem}
Define the spaces $\cW_{op}^{\mu,s} :=H(\cW^{\mu,s})$
and $\cW_{mn,op}^{\mu,s} :=H(\cW_{mn}^{\mu,s})$. Sometimes we
display the parameter $\xi$ as in $\cW_{op,\xi}^{\mu,s}
:=H(\cW^{\mu,s}_\xi)$. Theorem \ref{thm-III-1-2} implies that
$H(\cW^{\mu,s})$  is a Banach space under the norm $\big\| \, H(\uw)
\big\|_{\mu,s, \xi}$ $:=\ \big\| \, \uw \, \big\|_{\mu,s, \xi}$.
Similarly, the other spaces defined above are Banach spaces in the
corresponding norms.

Recall that $B$ denotes the dilation generator on the Fock space
$\cF$ (see \eqref{eq-I.24}). Let
\begin{equation}
\chi_\rho\equiv\chi_{H_f\le\rho}\ \mbox{and}\
\overline{\chi}_\rho\equiv\chi_{H_f\ge\rho}
\end{equation}
be a smooth partition of unity,
$\chi_\rho^2+\overline{\chi}_\rho^2=\one$. Let
$F_\rho:=F_{\tau\chi_\rho}$. We have
\begin{lemma} \label{LAPprepar}
Let $\chi_\rho^\#$ be either $\chi_\rho$ or $\overline{\chi}_\rho$.
If $H\in  \cW_{op}^{\mu,1} $, then the operators
\begin{equation}
\adj_B^j(\chi_\rho^\#),\ H_f^{-1}\adj_B^j(W_{00})\ \mbox{and}\
\adj_B^j(H-W_{00})\ \mbox{are bounded}
\end{equation}
for $j\leq 1$.  In particular, condition \eqref{eqn:5} with
$\tau(H):=W_{00}$, and therefore property \eqref{eqn:7}, with
$\chi=\chi_\rho$, hold for $H(\lambda)\in
C^\infty(\Delta,\cW_{op}^{\mu,1})\cap\dom(F_\rho)$.
\end{lemma}
\Proof  The result follows from the following relations
\begin{align}
\comm{B}{a^\#(k)}=\pm i( k\cdot\nabla_k+\frac{d}{2} ) a^\#(k),\\
i\comm{B}{H_f}=H_f, i\comm{B}{f(H_f)}=H_f f'(H_f).
\end{align}
Using these relations we show, in particular, that if $H\in
\cW_{op,\xi}^{\mu,1} $, then for any $j\leq 1,\ \adj_B^j(W)\in
\cW_{op,\xi'}^{\mu,1-j}\ \forall \xi' < \xi$ and
\begin{equation} \label{eq-III-1-26}
\|ad_B(H_{m n})\|_{\cW_{m n, op}^{\mu,0}} \leq c(m+n+1)\|H_{m
n}\|_{\cW_{m n, op}^{\mu,1}}.
\end{equation}
Eqn (III.25) implies that
$\adj_B^j(\chi_\rho^\#)$ and $H_f^{-1}\adj_B^j(T)$ are bounded and
Eqn \eqref{eq-III-1-26} implies that $\adj_B^j(W)$ are bounded, for
 $j\leq 1$.
\qed

%
\secct{The Renormalization Transformation $\cR_\rho$} \label{sec-IV}
In this section we present an operator-theoretic renormalization
transformation based on the smooth Feshbach-Schur map related
closely to the one defined in \cite{BachChenFroehlichSigal2003} and
\cite{BachFroehlichSigal1998a,BachFroehlichSigal1998b}. We fix the
index $\mu$ in our Banach spaces at some positive value $\mu > 0$.

The renormalization transformation is homothetic to an isospectral
map defined on a subset of a suitable Banach space of
Hamiltonians. It has a certain contraction property which insures
that (upon an appropriate tuning of the spectral parameter) its
iteration converges to a fixed-point (limiting) Hamiltonian, whose
spectral analysis is particularly simple. Thanks to the
isospectrality of the renormalization map, certain properties of
the spectrum of the initial Hamiltonian can be studied by
analyzing the limiting Hamiltonian.

The renormalization map is defined below as  a composition of a
decimation map, $F_{\rho}$, and two rescaling maps, $S_\rho$ and
$A_\rho$. Here $\rho$ is a positive parameter - the photon energy
scale - which will be chosen later.


The \emph{decimation of degrees of freedom} is done by the smooth
Feshbach map, $F_{\tau,\chi}$. Except for the first step, the
decimation map  will act on the Banach space $\cW_{op}^s$. The
operators $\tau$ and $\chi$ will be chosen as
\begin{equation}
\tau(H)=W_{00}:=w_{0 0}(H_f)\ \mbox{and}\
\chi=\chi_\rho\equiv\chi_{\rho^{-1}H_f\le1} , \label{IV.1}
\end{equation}
where $H=H(\uw)$ is given in  Eqn \eqref{III.12}. With $\tau$ and
$\chi$ identified in this way we will use the notation
\begin{equation}
F_\rho\equiv F_{\tau,\chi_\rho}. \label{IV.2}
\end{equation}
The following lemma shows that the domain of this map contains the
following polydisc in $\cW_{op}^{\mu,s}$:
\begin{eqnarray}
\cD^{\mu,s}(\alpha,\beta,\gamma) & := & \Big\{ H(\uw) \in
\cW_{op}^{\mu,s} \ \Big| \
 | w_{0,0}[0]| \leq \alpha \comma
\\ \nonumber & &
 \sup_{r \in [0,\infty)}| w_{0,0}'[r] - 1 | \leq \beta
 \comma \hspace{4mm}
\| \uw_1 \|_{\mu,s, \xi} \leq \gamma \Big\} \comma
\label{IV-3}\end{eqnarray}
for appropriate $\alpha, \beta, \gamma >0$. Here
$\uw_1 :=(w_{m,n})_{m+n \geq 1}$.

\begin{lemma} \label{lem-III-2-2}
Fix $0 < \rho < 1$, $\mu > 0$, and $0 < \xi < 1$. Then it
follows that the polidisc $\cD^{\mu,1}(\rho/8, 1/8, \rho/8)$ is in
the domain of the Feshbach map $F_\rho$.
\end{lemma}
\Proof Let  $H(\uw) \in \cD^{\mu,1}(\rho/8, 1/8, \rho/8)$. We remark
that $W:= H[\uw]-W_{0,0}[\uw]$ defines a bounded operator on $\cF$,
and we only need to check the invertibility of $H(\uw)_{\tau
\chi_\rho}$ on $\Ran \,\bchi_\rho$. Now the operator $T + E =
W_{0,0}[\uw]$ is invertible on $\Ran \,\bchi_\rho$ since
for all $r \in [3\rho/4, \infty)$
\begin{eqnarray} \label{eq-III-2-16}
\rRe w_{0,0}[r] & \geq & r \, - \, | w_{0,0}[r] - r | \nonumber
\\ & \geq & r \big( 1 \, - \, \sup_{r} | w_{0,0}'[r] - 1 | \big) \: - \:  |w_{0,0}[0]| \nonumber \\
& \geq & \frac{3 \, \rho}{4} ( 1 - 1/8 ) \: - \: \frac{\rho}{8} \
\geq \ \frac{ \rho}{2} \ .
\end{eqnarray}
%
On the other hand, by \eqref{III.11}, $\big\| W \|\leq \xi\rho/8
\leq \rho/8$. Hence $\rRe (W_{0,0}[\uw] + W) \geq \frac{\rho}{3}$ on
$\Ran \,\bchi_\rho$, i.e. $H(\uw)_{\tau, \bchi_\rho}$ is invertible
on $\Ran \,\bchi_\rho$. \QED

We introduce the \textit{scaling transformation} $S_\rho: \cB[\cF]
\to \cB[\cF]$, by
%
%
%
\begin{equation} \label{IV.5}
S_\rho(\one) \ := \ \one \comma \hspace{5mm} S_\rho (a^\#(k)) := \
\rho^{-d/2} \, a^\#( \rho^{-1} k) \comma
\end{equation}
where $a^\#(k)$ is either $a(k)$ or $a^*(k)$, and $k \in \RR^d$.
%
On the domain of the decimation map $F_\rho$ we define   the
renormalization map $\cR_\rho$ as
\begin{equation} \label{IV.6}
\cR_\rho:=  \rho^{-1} S_\rho\circ F_\rho.
\end{equation}

\begin{remark} \label{remIV-2} The renormalization map above is different from the one
defined in \cite{BachChenFroehlichSigal2003}.
The map in \cite{BachChenFroehlichSigal2003}  contains an additional
change of the spectral parameter $\lambda:= -\la H\ra_\Omega$.
\end{remark}

We mention here some properties of the scaling transformation. It is
easy to check that $S_\rho (\hf) = \rho \hf$, and hence
\begin{equation} \label{eq-III-2-3}
S_\rho ( \chi_\rho) = \ \chi_1 \hspace{5mm} \mbox{and}
\hspace{6mm} \rho^{-1} S_\rho \big( \hf \big) \ = \  \hf
\comma
\end{equation}
%
%
%
which means that the operator $\hf$ is a \emph{fixed point} of
$\rho^{-1} S_\rho$. Further note that $E \cdot \one$ \emph{is
expanded} under the scaling map, $\rho^{-1} S_\rho(E \cdot
\one) = \rho^{-1} E \cdot \one$, at a rate $\rho^{-1}$. (To
control this expansion it is necessary to suitably restrict the
spectral parameter.)

Now we show that the interaction $W$ contracts under the scaling
transformation. To this end we remark that the scaling map $S_\rho$
restricted to $\cW_{op}^{\mu,s}$
induces a scaling map $s_\rho$ on $\cW^{\mu,s}$ by
\begin{equation} \label{eq-III-2-5}
\rho^{-1} S_\rho \big( H(\uw) \big) \ =: \ H \big( s_\rho(\uw)
\big)  \comma
\end{equation}
where $s_\rho(\uw):=(s_\rho(w_{m,n}) )_{m+n \geq 0} $, and it is
easy to verify that, for all $(m,n) \in \NN_0^2$,
\begin{equation} \label{eq-III-2-6}
s_\rho(w_{m,n}) \big[ r , k_{(m,n)} \big] \ = \ \rho^{m+n - 1} \:
w_{m,n}\big[ \rho \, r \; , \; \rho \, k_{(m,n)} \big] \period
\end{equation}
%
%
%
We note that by Theorem~\ref{thm-III.1}, the operator norm of
$W_{m,n} \big[ s_\rho(w_{m,n}) \big]$ is controlled by the norm
$$ \| s_\rho(w_{m,n}) \|_{\mu} =$$
\begin{eqnarray}\nonumber
& \max_j &\sup_{r \in I, k \in B_1^{m+n}} \ \rho^{m+n - 1} \:
\frac{\big|w_{m,n}[\rho \, r \; , \; \rho \, k_{(m,n)}] \big|}{|
k_j|^{\mu}}
%
\\ \nonumber & \leq &
\rho^{m+n +\mu- 1} \, \| w_{m,n} \|_{\mu}.
\end{eqnarray}
Hence, for $m+n \geq 1$, we have
\begin{equation} \label{eq-III-2-8}
 \| s_\rho(w_{m,n}) \|_{\mu}  \leq \ \; \rho^{\mu}
\, \| w_{m,n} \|_{\mu}
\end{equation}
Since $\mu >0$, this estimate shows that $S_\rho$ contracts $\|
w_{m,n} \|_{\mu}$ by at least a factor of $\rho^{\mu} < 1$. The next
result shows that this contraction is actually a dominating property
of the renormalization map $\cR_\rho$ along the 'stable' directions.
Below, recall, $\chi_{1}$ is the cut-off function introduced at the
beginning of Section III. Define the constant
\begin{equation} \label{VIII.21}
C_\chi:=\frac{4}{3}\big(\sum_{n=0}^2 \sup |
\partial_r^n \chi_1| + \sup |\partial_r \chi_1 |^2 \big) \le 200.
\end{equation}
%
\begin{theorem}
\label{thm-III-2-5} Let $\epsilon_0:H\rightarrow \la H\ra_\Omega$
and $\mu>0$ (see \eqref{IV-3}). Then
for the absolute constant $C_\chi$
given in \eqref{VIII.21} and for any $s \ge 1,\ 0<\rho< 1/2,
\alpha,\beta \le \frac{\rho}{8}$ and $\gamma \le
\frac{\rho}{8C_\chi}$ we have
\begin{equation}
\cR_\rho-\rho^{-1}\epsilon_0:\cD^{\mu,s}(\alpha,\beta,\gamma)\rightarrow
\cD^{\mu,s}(\alpha',\beta',\gamma'), \label{eqn:23}
\end{equation}
continuously, with $\xi:=\frac{\sqrt{\rho}}{4C_\chi}$ (in the
definition of the corresponding norms) and
\begin{equation}
\alpha'=3C_\chi\lb\gamma^2/2\rho\rb,
\beta'=\beta+3C_\chi\lb\gamma^2/2\rho\rb,
\gamma'=128C_\chi^2\rho^\mu\gamma \label{eqn:24}
\end{equation}
\end{theorem}
With some modifications, this theorem follows from
\cite{BachChenFroehlichSigal2003}, Theorem 3.8 and its proof,
especially Equations (3.104), (3.107) and (3.109). For the norms
\eqref{III.5} with $q=0$ it is presented in \cite{Sigal}, Appendix
I. A generalization to the $q>0$ case is straightforward.

\begin{remark} \label{remIV-4} Subtracting the term $\rho^{-1}\epsilon_0$ from $\cR_\rho$
allows us to control the expanding direction during the iteration of
the map $\cR_\rho$. In \cite{BachChenFroehlichSigal2003} such a
control was achieved by using the change of the spectral parameter
$\lambda$ which controls $\la H\ra_\Omega$ (see remark in Appendix
I).
%
\end{remark}

\begin{proposition} \label{prop-LAP}
Let $\Delta$ be an open interval in $\mathbb{R}$, $\mu>0$ and let $\rho$ and $\xi$ be as in Theorem
\ref{thm-III-2-5}. Then for $H(\lambda)\in C^\infty(\Delta,
\cD^{\mu,1}(\alpha,\beta,\gamma))$, with
$\alpha,\gamma<\frac{\rho}{8}, \beta \le \frac{1}{8}$, the following
is true for $1 \ge \theta > 0$ and $\nu \ge 0$

\begin{equation}
B_\theta(\cR_\rho(H(\lambda))-i0)^{-1}B_\theta\in
C^\nu(\Delta)\Rightarrow B_\theta(H(\lambda)-i 0)^{-1}B_\theta\in
C^\nu(\Delta) \label{LAP}.
\end{equation}
\end{proposition}

\Proof By Theorem
\ref{thm-III-2-5}, $\forall \lambda \in \Delta, H(\lambda) \in \dom
\big(\cR_\rho\big)$. Then Lemma \ref{LAPprepar}
and invariance of the operator $B_\theta$ under the rescaling
$S_\rho$ imply the result. \qed

\secct{Renormalization Group} \label{sec-RG}
%
In this section we describe some dynamical properties of the
renormalization group $\cR_\rho^n\ \forall n \ge 1 $  generated by
the renormalization map $\cR_\rho$. A closely related iteration
scheme is used in \cite{BachChenFroehlichSigal2003}. First, we
observe that $\forall w \in \mathbb{C}, \cR_\rho(wH_f) = wH_f$ and $
\cR_\rho(w\textbf{1}) =\frac{1}{\rho} w\textbf{1}$. Hence we define
$\cM_{fp}:=\mathbb{C}H_f$  and $\cM_{u}:=\mathbb{C}\textbf{1}$ as
candidates for  a manifold of the fixed points of $\cR_\rho$ and an
unstable manifold for $\cM_{fp}:=\mathbb{C}H_f$. The next theorem
identifies the stable manifold of $\cM_{fp}$ which turns out to be
of the (complex) codimension $\one$ and is foliated by the (complex)
co-dimension $2$ stable manifolds for each fixed point in
$\cM_{fp}$. This implies in particular that in a vicinity of
$\cM_{fp}$ there are no other fixed points and that $\cM_{u}$ is the
entire unstable manifold of $\cM_{fp}$.

We introduce some definitions.
As an initial set of operators we take
$\cD:=\cD^{\mu,2}(\alpha_0,\beta_0,\gamma_0)$ with $\alpha_0,
\beta_0,\gamma_0 \ll 1$. (The choice $s=2$ of the smoothness index
in the definition of the polidiscs is dictated by the needs of the
Mourre theory applied in the next section.) We also let
$\cD_s:=\cD^{\mu,2}(0,\beta_0,\gamma_0)$ (the subindex s stands for
'stable', not to be confused with the smoothness index $s$ which in
this section is taken to be 2). We fix the scale $\rho$ so that
\begin{equation}
\alpha_0, \beta_0,\gamma_0 \ll\rho\le min(\frac{1}{2}, C_\chi^2)
\label{rho}
\end{equation}
where, recall,  the constant $C_\chi$ is appears in Theorem
\ref{thm-III-2-5} and is defined in \eqref{VIII.21}. Below we will
use the $n-$th iteration of the numbers $\alpha_0, \beta_0$ and
$\gamma_0$ under the map \eqref{eqn:24}:
$$\alpha_n:=c\lb\rho^{-1}(c\rho^\mu)^{n-1}\gamma_0\rb^2,$$
\begin{equation*}
\beta_n=\beta_0+\sum_{j=1}^{n-1}
c\lb\rho^{-1}(c\rho^\mu)^j\gamma_0\rho\rb^2,
\end{equation*}
$$\gamma_n=(c\rho^\mu)^n\gamma_0.$$
For $H \in \cD$ we denote $H_u:=\la H\ra_\Omega$ and $H_s:=H - \la
H\ra_\Omega\ \mathbf{1}$ (the unstable- and stable-central-space
components of $H$, respectively). Note that $H_s \in \cD_s$.


Recall that a complex function $f$ on an open set $\cD$ in a complex
Banach space $\cW$ is said to be \textit{analytic} if $\forall \xi
\in \cW,\ f(H+ \tau \xi)$ is analytic in the complex variable $\tau$
for $|\tau|$ sufficiently small (see \cite{Berger}). Our analysis
uses the following result from \cite{FroehlichGriesemerSigal2009}:
\begin{theorem} \label{stable-manif} Let $\delta_n :=\nu_n\rho^{n}$ with
$4 \alpha_n \leq \nu_{n} \leq\frac{1}{18}$. There is
an analytic  map $e:\cD_s \rightarrow \mathbb{C}$ s.t. $e(H) \in
\mathbb{R}$ for $H=H^*$ and
\begin{equation}
U_{\delta_n} \subset dom(\cR_\rho^{n})\ \mbox{and}\
\cR_\rho^{n}(U_{\delta_n}) \subset
\cD^{\mu,2}(\rho/8,\beta_{n},\gamma_{n}) \label{eqn:30aaa}
\end{equation}
where $U_\delta:= \{H \in \cD|\ |e(H_s)+ H_u|  \leq \delta\ \}.$
Moreover, $\forall H \in U_{\delta_n}$ and $\forall n \geq 1$, there
are $E_{n} \in \mathbb{C}$ and $w_n(r)\in \mathbb{C}$ s.t. $|E_{n}|
\leq 2\nu_n$, $|w_n(r)-1| \leq \beta_n$,  $ w_n$ is $C^2$,
\begin{equation}
\cR_\rho^n(H)=E_{n}+w_n(H_f)H_f +O_{\cW_{op}^s}(\gamma_n),
\label{eqn:30b}
\end{equation}
$E_{n}$ and $w_n(r)$ are real if $H$ is self-adjoint and, as $ n
\rightarrow \infty$.
\end{theorem}
Moreover, one can show that   $w_n(r) $ converge in $L^\infty$ to
some number (constant function) $w\in \mathbb{C}$
(\cite{FroehlichGriesemerSigal2009}).

This theorem implies that $\cM_{fp}:=\mathbb{C}H_f$  is (locally) a
manifold of the fixed points of $\cR_\rho$ and
$\cM_{u}:=\mathbb{C}\textbf{1}$ is an unstable manifold
and the set
\begin{equation} \cM_s:=\bigcap_n U_{\delta_n}= \{H\in \cD |\ e(H_s)=-H_u\}\label{eqn:30a}
\end{equation}
is a local stable manifold for the fixed point manifold $\cM_{fp}$
in the sense that $\forall H \in \cM_s\ \exists w \in \mathbb{C}$
s.t.
\begin{equation}\cR_\rho^n(H)\rightarrow wH_f\ \mbox{in the sense of}\
 \cW_{op}^s  \label{eqn:30aa}
\end{equation}
as $ n \rightarrow \infty$. Moreover, $\cM_s$ is an invariant
manifold for $\cR_\rho$: $\cM_s \subset \dom(\cR_\rho)$ and
$\cR_\rho(\cM_s) \subset \cM_s$, though we do not need this property
here and therefore we do not show it.
The next result reveals the spectral significance of the map $e$:
\begin{theorem}
Let $H\in \cD$. Then the number $E:=e(H_s)+H_u$ is an eigenvalue of
the operator $H$. Moreover, if $H$ is self-adjoint, then it is the
ground state energy of $H$.
\end{theorem}

Theorems V.2 and V.3 were proven in \cite{Sigal} for somewhat simper
Banach spaces which do not contain the derivatives
$(k\partial_k)^q$). However, an extension to the Banach spaces which
are used in this paper is straightforward and is omitted here.
\secct{Mourre Estimate} \label{sec-VII}
%
In this section we prove
the Mourre estimate for the operator-family $H^{(n)}(\lambda):=
\cR_{\rho}^n(H)$ with $\lambda:=-H_u$.
This gives the limiting absorption principle for $H^{(n)}(\lambda)$.
The latter is then transferred with the help of Theorem \ref{LAPtransfer} to the
limiting absorption principle for the operator $H$. In Section \ref{sec-VIII}
this limiting absorption principle will be connected to the limiting
absorption principle for the family $H_g-\lambda$, where $H_g$ is
either $H^{PF}_g$ or $H^{N}_g$.

\begin{theorem}
\label{thm-VII-1} Let $H(\lambda)=H(\lambda)^*\in
C^\infty(\Delta,\cD^{\mu,2}(\alpha,\beta,\gamma)),$ where $\Delta$ is an open interval in $\mathbb{R}$,  and
$\Delta^\delta:=[\delta,\infty)$.  If $\delta\gg \gamma$ and $\beta
\leq \frac{1}{3}$, then
\begin{equation}
B_\theta(H(\lambda)-i0)^{-1}B_\theta\in C^\nu(\Delta\cap
E^{-1}(\Delta^\delta)), \label{eqn:67}
\end{equation}
where $E:\lambda\rightarrow E(\lambda)$ with $E(\lambda):=\la
H(\lambda)\ra_\Omega$, for any  and $1/2<\theta \le 1$ and
$\nu<\theta-\frac{1}{2}$.
\end{theorem}

\Proof In what follows we omit the argument $\lambda$.  Let
$E:=w_{0,0}[0], T:=w_{0,0}[\hf] - w_{0,0}[0]\ \mbox{and}\
W:=\sum_{m+n \geq 1}\chi_1 W_{m,n}[\uw]\chi_1$, so that $H=E\one + T
+ W$. Let
$H_1:=H-E=T+W$. We write $i \comm{H_1}{B}=\tT+\tW$, where $\tT:=i
\comm{T}{B}=T'(H_f)H_f$ and $\tW:=i \comm{W}{B}$. By relation
(III.28) we have for $s=2$
$$\|\tW\|_{\cW_{op}^{\mu,s-1}}\le c\gamma,$$
where the $\xi$-parameter in the norm on the l.h.s. should be taken
slightly smaller than the $\xi$-parameter in the Banach space
$\cW_{op}^{\mu,s}$ for $W$. The shift in the smoothness index from
$s$ to $s-1$ is due to the fact that the coupling functions for the
operator $i \comm{W}{B}$ are $(k \cdot \nabla_k +
\frac{3(m+n)}{2})w_{m,n}(r,k)$, where $k:=k^{(m,n)}$, and therefore
loose one derivative compared to the coupling functions,
$w_{m,n}(r,k)$, of
 $W$.

We write
\begin{equation*}
i\comm{H_1}{B}=\frac{1}{2}H_1+\tT-\frac{1}{2} T+\tW-\frac{1}{2}W.
\end{equation*}
Remembering that the operator norm is dominated by the
$\cW^{\mu,0}_{op}-$ norm we see that the last two terms are bounded
as
\begin{equation}
\|\tW-\frac{1}{2}W\|\le C\gamma. \label{eqn:69}
\end{equation}
Furthermore using the estimate $|T'(r) - 1| < \beta$ and the
definition of $\tT$ we find
%
\begin{equation*}
\tT(r)-\frac{1}{2}T(r)\ge
(1-\beta)r-\frac{1}{2}(1+\beta)r=\frac{1}{2}(1-3\beta)r
\end{equation*}
and therefore
$$\tT-\frac{1}{2}T\ge\inf_{0\le r\le \infty}\lb \tT(r)-\frac{1}{2}T(r)
\rb $$
\begin{equation*}
\ge\inf_{0\le r\le \infty}\frac{1}{2}(1-3\beta)r=0.\label{eqn:70}
\end{equation*}
This gives
$\comm{H_1}{B}\ge\frac{1}{2}H_1-c\gamma$
and therefore for $\Delta':=\lb\frac{1}{2}\delta,\infty\rb$,
$\delta\gg\gamma$,
\begin{equation}
E_{\Delta'}(H_1)i\comm{H_1}{B}E_{\Delta'}(H_1)\ge\frac{1}{4}\delta
E_{\Delta'}(H_1)^2. \label{eqn:71}
\end{equation}
This proves the Mourre estimate for the operator $H_1\equiv
H_1(\lambda)$.

Moreover, since $H(\lambda)\in
C^\infty(\Delta,\cD^{\mu,2}(\alpha,\beta,\gamma))$, we have that the
commutators $[H_1, B]$ and $[[H_1, B], B]$ are bouded relative to
the operator $H_1$ (this is guaranteed by taking the index $s=2$ for
the polidisc $D^{\mu,s}(\alpha,\beta,\gamma)$). Hence the standard
Mourre theory is applicable and
gives  H\"{o}lder continuity in the spectral parameter $\sigma$ as
well as in the "operator $H_1(\lambda)$", i.e. in $\lambda$ (see
\cite{HunzikerSigal}):
\begin{equation}
B_\theta R_1(\lambda,\sigma)E_{\Delta'}(H_1(\lambda))B_\theta\in
C^\nu(\Delta\times\RR), \label{eqn:72}
\end{equation}
where $\nu < \theta -1/2$, where we restored the argument $\lambda$
in our notation and where
$R_1(\lambda,\sigma):=(H_1(\lambda)-\sigma)^{-1}$.  Since
$$B_\theta R_1(\lambda,\sigma)B_\theta=B_\theta
R_1(\lambda,\sigma)E_{\Delta'}(H_1(\lambda))B_\theta$$
\begin{equation}
+B_\theta
R_1(\lambda,\sigma)(\one-E_{\Delta'}(H_1(\lambda)))B_\theta
\label{eqn:73}
\end{equation}
and since the last term on the right hand side is $C^\nu(\Delta)$ in
$\lambda$ and $C^\infty(\Delta^\delta)$ in $\sigma$ we conclude from
\eqref{eqn:72} that
\begin{equation}
B_\theta R_1(\lambda,\sigma) B_\theta\in
C^\nu(\Delta\times\Delta^\delta). \label{eqn:74}
\end{equation}
Now take $\sigma = E(\lambda)+i0$. Since by the condition of the
theorem $E(\lambda):=\la H(\lambda)\ra_\Omega\in C^\infty(\Delta)$
we conclude that \eqref{eqn:67} holds.\qed

In the previous section the parameter $\delta_n$ was allowed to
change in a certain range (see Theorem V.2). In this section we make
a particular choice of $\delta_n$, namely $\delta_n
:=\frac{1}{18}\rho^{n}$. Recall the definition of the set $U_{\delta}$ in Theorem \ref{stable-manif}.
\begin{theorem} \label{thm-VII-2}
Assume \eqref{rho}.
Let $ n \geq 1$, $\delta_n
:=\frac{1}{18}\rho^{n}$ and let $ H=H^* \in U_{\delta_n}
$
and $\Delta_{\delta_n}:=
[e(H_s)+\frac{\rho}{2}\delta_n,e(H_s)+\delta_n]$. Then
\begin{equation}
B_\theta(H_s-\lambda-i0)^{-1}B_\theta\in C^\nu(\Delta_{\delta_n})
\label{eqn:75}
\end{equation}
for any  and $1/2<\theta \le 1$ and $\nu<\theta-\frac{1}{2}$.
\end{theorem}
\Proof Let $D_n$ be the disc of the radius
$\delta_n$ centered at $e(H_s)$. Since, by \eqref{eqn:30aaa},  $U_{\delta_n}
\subset D(\cR_\rho^{n})$,
the operator $H^{(n)}(\lambda):=\cR_\rho^{n}(H)$, with
$\lambda:=-H_u$, is well defined.  By \eqref{eqn:30aaa},
%
$D_n\ni\lambda\rightarrow H^{(n)}(\lambda)\in
\cD^{\mu,2}(\frac{1}{8}\rho,\beta_{n-1},\gamma_{n-1})$ is
$C^\infty$. Moreover, $H^{(n)}(\lambda)=H^{(n)}(\lambda)^*,\ \forall \lambda \in D_n \cap \mathbb{R}$. Hence, since $\rho\gg\gamma_{n-1}$, $\beta_{n-1} \leq
\frac{1}{3}$, by \eqref{rho}, and $\Delta_{\delta_n} \subset D_n$,
we have by Theorem \ref{thm-VII-1} that
\begin{equation*}
B_\theta(H^{(n)}(\lambda)-i0)^{-1} B_\theta\in C^\nu(D_n\cap
E_n^{-1}(\Delta^{\frac{1}{50}\rho})),
\end{equation*}
where $0 \le \nu < \theta - 1/2$ and, as before,
$E_{n}(\lambda)\equiv E_{n}(\lambda, H_s):=\big(
H^{(n)}(\lambda)\big)_u$,
which, by the above conclusion, is $C^\infty$.
We need the following proposition to describe the set $E_n^{-1}(\Delta^{\frac{1}{50}\rho})$.
\begin{proposition}
\label{prop-VII-3} Let $n\geq0$,
$\delta_n :=\frac{1}{18}\rho^{n}$ and $A_{\delta_n}:=
\{\frac{\rho}{2}\delta_n \leq |\lambda -e(H_s)| \leq \delta_n \}$.
For $H \in U_{\delta_n} $ we denote $E_{n}(\lambda, H_s):=\big(
\cR_{\rho}^n(H)\big)_u\equiv\la \cR_{\rho}^n(H)\ra_\Omega,\ \lambda
=-H_u$. Then
\begin{equation}
   |E_{n}(\lambda, H_s)|\ge  \frac{1}{50}\rho\ \,\
    \mbox{for}\ \lambda\in A_{\delta_n}.
    \label{eqn:63}
\end{equation}
\end{proposition}
\Proof In this proof we do not display the argument $H_s$. Let
$\lambda \in A_{\delta_n}$ with $\delta_n$ given in the proposition.
Define $E_{0 i}(\lambda)$ by the equation
\begin{equation}
E_{n}(\lambda)=\rho^{-n}(E_{0 n}(\lambda)-\lambda). \label{eqn:55}
\end{equation}
The following estimate is shown in \cite{Sigal} (see Eqn (V.27) of
the latter paper):
%
%
%
%
%
%
\begin{equation}
|E_{0 n}(\lambda)-e| \le\frac{1}{5}|\lambda-e|+
(1-\rho)^{-1}\rho^{n+1} \alpha_{n+1}.
\end{equation}
This inequality and the definition of $\alpha_n$ imply
\begin{align}
|E_{0
n}(\lambda)-\lambda|&\ge|\lambda-e|-|E_{0 n}(\lambda)-e|\nonumber\\
&\ge \frac{4}{5}|\lambda-e|-2\gamma_0^2
c(c^2\rho^{2\mu+1})^n\rho^{-1}.\label{eqn:64}
\end{align}
Due to
$2\gamma_0^2 c\rho^{-1}(c^2\rho^{2\mu+1})^{n}\ll \rho^{n},$
\eqref{eqn:64} gives
\begin{equation}
 |E_{0 n}(\lambda)-\lambda| \ge \frac{1}{50}\rho^{n+1}. \label{eqn:65}
\end{equation}
Due to \eqref{eqn:55}
this implies
the statement of the proposition.\qed
%

Proposition \ref{prop-VII-3} says that
\begin{equation}
E_n:\Delta_{\delta_n}\ni\lambda\rightarrow E_{n}(\lambda)\in\
\Delta^{\frac{1}{50}\rho}.
\end{equation}
Hence $E_{n}^{-1}(\Delta^{\frac{1}{50}\rho}) \supset
\Delta_{\delta_n} $. Since $\Delta_{\delta_n} \subset D_n$, we have that
\begin{equation}
B_\theta(H^{(n)}(\lambda)-i0)^{-1}B_\theta\in
C^\alpha(\Delta_{\delta_n}) \label{eqn:75a}
\end{equation}
which, due to Proposition  \ref{prop-LAP}, gives \eqref{eqn:75}.
\qed
%

%
\secct{Initial Conditions for the Renormalization Group}
\label{sec-IC}

Now we turn to the operator families $H_g-\lambda$, we are
interested in. Here the operator $H_g=H_0+gI$ is given either by
~\eqref{Hn} or by  (~\ref{Hpf}). These operators do not belong to
the Banach spaces defined above. We define an additional
renormalization transformation which acts on such operators and maps
them into the disc $\cD^{\mu,s}(\alpha_0,\beta_0,\gamma_0)$ for some
appropriate $\alpha_0$, $\beta_0$, $\gamma_0$.

Let  $H_{pg}$ denote either  $H^{PF}_{p}$ or  $H^{N}_{p}$ and let
$e^{(p)}_{0}< e^{(p)}_{1} < ...$ be the eigenvalues of $H_{pg}$, 
so that $e^{(p)}_{0}$ is its the ground state
energy. 
Let $P_{p}$ be the orthogonal projection onto the eigenspace
corresponding to $e^{(p)}_0$. On Hamiltonians acting on
$\cH_{p}\otimes\cH_f$ which were described above, we define the map
\begin{equation}
\cR_{\rho_0}^{(0)}=\rho_0^{-1} S_{\rho_0}\circ F_{\tau_0 \pi_0},
\label{eqn:25}
\end{equation}
where $ \rho_0 \in (0, \e^{(p)}_{gap}]$  is an initial photon energy
scale (recall that $\e^{(p)}_{gap}:= \e^{(p)}_1-\e^{(p)}_0$ and $\e^{(p)}_j$ are the eigenvalues of $H_p$) and
where
\begin{equation}
\tau_0(H_g-\lambda)=H_{0g}-\lambda\ \mbox{and}\ \pi_0 \equiv
\pi_0[H_f]:=P_{p}\otimes\chi_{H_f\le\rho_0}. \label{pi_0}
\end{equation}
for any $\lambda \in \mathbb{C}$. Recall the convention $\bar{\pi}_0
:= \textbf{1} - \pi_0$. Define the set
\begin{equation}I_0:=\{z\in \mathbb{C}|Rez \leq e^{(p)}_0 +
\frac{1}{2}\rho_0\}.\label{eqn:26a}
\end{equation}

We assume  $\rho_0\gg g^2$.
%
To simplify the notation we assume that the ground state energy,
$\e^{(p)}_0$, of the operator $H_{p}$ is simple (otherwise we would
have to deal with matrix-valued operators on $\cH_f$).  We have
\begin{theorem}
\label{thm-V.1} Let $H_g$ be the Hamiltonian given either by
~\eqref{Hpf} or by   ~\eqref{Hn} and let $\rho_0\gg g^2,\ \mu >
-1/2$ and $\lambda\in I_0$.
Then
\begin{equation}
H_g-\lambda\in \dom(\cR_{\rho_0}^{(0)}). \label{V.4}
\end{equation}
Furthermore, define the family of operators
$H_{\lambda}^{(0)}:=\cR_{\rho_0}^{(0)}(H_g -\lambda) \mid \Ran
{P}_{p}\otimes\ \textbf{1}$. Then $H_{\lambda}^{(0)}=H_{\lambda}^{(0)*}$, for $\lambda\in I_0\bigcap \mathbb{R} $, and  
\begin{equation}
H_{\lambda}^{(0)}-\rho_0^{-1}(e^{(p)}_{0}-\lambda)\in
\cD^{\mu,2}(\alpha_0,\beta_0,\gamma_0), \label{V.5}
\end{equation}
where, with $\mu$ as in Eqn ~\eqref{I.7}, $\alpha_0=O(g^2
\rho_0^{-1})$, $\beta_0=O(g^2)$, and $\gamma_0=O(g\rho_0^\mu)$, for  $\lambda\in I_0$.
Moreover, $\cR_{\rho_0}^{(0)}(H-\lambda)$ is analytic in $\lambda\in
I_0$. In particular, these results apply to the Pauli-Fierz and
Nelson Hamiltonians by taking $\mu=1/2$ and $\mu>0$, respectively.
\end{theorem}
Note that if $\psi^{(p)}$ is a ground state of $H_{pg}$ with the
energy $e^{(p)}_{0}$ and $\psi_0=\psi^{(p)}\otimes\Omega$, then we
have
\begin{equation}
e^{(p)}_{0}-\lambda=\la H-\lambda\ra_{\psi_0}.
\end{equation}

Theorem \ref{thm-V.1} is proven in \cite{Sigal}, Appendix II, for
somewhat simper Banach spaces which do not contain the derivatives
$(k\partial_k)^q$. However, an extension to the Banach spaces which
are used in this paper is straightforward and is omitted here.


Note that $ K:= \cR_{\rho_0}^{(0)}(H_g-\lambda)\mid _{\Ran
({\bar{P}}_{pj}\otimes\ \mathbf{1})}= (H_{0g}-\lambda)\mid_{ \Ran
({\bar{P}}_{pj}\otimes\ \mathbf{1})} $ and therefore $\forall \lambda \in I_0 \cap \mathbb{R},\ \sigma(K) =
\sigma(H_{pg})/\{\lambda_j\} +[0, \infty) - \lambda$. Hence
$$\forall \lambda \in I_0 \cap \mathbb{R}, K \ge e^{(p)}_1-e^{(p)}_0 -
\frac{1}{8}\rho_0\ge \frac{7}{8}(e^{(p)}_1-e^{(p)}_0).$$ Therefore
$0 \notin \sigma(K)$.  This, the relation $\sigma
(\cR_{\rho_0}^{(0)}(H_g-\lambda)) = \sigma(H_{\lambda}^{(0)}) \cup
\sigma(K)$ and Theorem \ref{thm-II-1} imply
%
%
that $H_{\lambda}^{(0)}$ is isospectral to $H_g-\lambda$ in the
sense of Theorem \ref{thm-II-1}. Moreover, similarly to Proposition
IV.3, and using the relation
\begin{equation}
\cR_{\rho_0}^{(0)}(H_g-\lambda)^{-1}=H_{\lambda}^{(0)-1}(P_{pj}\otimes\
\mathbf{1}) +(H_{0g}-\lambda)^{-1}({\bar{P}}_{pj}\otimes\
\mathbf{1}). \label{VIII.7a}
\end{equation}
one shows the following result
\begin{proposition} Let $\mu>0$, $\rho_0 \gg g^2$ and
$\Delta_0\subseteq I_0\bigcap \mathbb{R} $.
If $H_g$ is given in either ~\eqref{Hpf} or ~\eqref{Hn}, then
\begin{equation} \label{V.19}
B_s(H_{\lambda}^{(0)}-i0)^{-1}B_s\in C^\nu(\Delta_0)\Rightarrow
B_s(H_g-\lambda-i0)^{-1}B_s\in C^\nu(\Delta_0).
\end{equation}
\end{proposition}
%

%
\secct{Proof of Theorem I.1} \label{sec-Thm I.1}

Let $H_g$ be a Hamiltonian
given in either ~\eqref{Hpf} or ~\eqref{Hn}. Recall the definition
\begin{equation}
\label{eqn:31} H^{(0)}_\mu:=\cR_{\rho_0}^{(0)}(H_g-\mu)\mid_{  \Ran
{P}_{p}\otimes\ \textbf{1}},\ \mu\in I_0.
\end{equation}
The r.h.s. is well defined according to Theorem~\ref{thm-V.1}.
By Equation \eqref{V.5}, if $\mu\in I_0$, then
\begin{equation}
H^{(0)}_\mu-\rho_0^{-1}(e^{(p)}_0-\mu)\in
\cD^{\mu,2}(\alpha_0,\beta_0,\gamma_0), \label{eqn:35}
\end{equation}
where $\alpha_0,\beta_0$ and $\gamma_0$ are given in Theorem
\ref{thm-V.1}. The condition \eqref{rho} is satisfied if
\begin{equation}
g^2\rho_0^{-1},\ g\rho_0^\mu\ll\rho\le \frac{1}{2},
\label{eqn:32}
\end{equation}
which can be arranged since by our assumption $g \ll 1$ and $\rho_0$
can be fixed anywhere in the interval $(0, \e^{(p)}_{gap}]$.

Let $H_{\mu s} : = (H^{(0)}_{\mu})_s=H^{(0)}_{\mu} - \la
H^{(0)}_{\mu}\ra_\Omega\ \mathbf{1}$ and $H_{\mu u} : =
(H^{(0)}_{\mu})_u = \la H^{(0)}_{\mu}\ra_\Omega $, the
stable-central and unstable components of the operator
$H^{(0)}_{\mu}$, respectively (see Section \ref{sec-RG}), and let
$e:\cD_s \rightarrow \mathbb{C}$ be the map introduced in Theorem
\ref{stable-manif}.
We introduce the subsets:
\begin{equation} \label{VII.20}
D_{\delta}: = \{ \mu \in I_0 | |e (H_{\mu s}) + H_{\mu u} | \le
\delta \}
\end{equation} and
\begin{equation} \label{VII.21} E^{\mu}_{\delta} : = \{ \lambda \in \mathbb{R}\
|\ {\rho \over 8} \delta \le | \lambda - e (H_{\mu s}) | \le \delta
\}.\end{equation} Recall, $\delta_n = {1 \over 18} \rho^n$ for $n
\ge 0$. Let $\theta
> {1 \over 2}$ and $0 < \nu < \theta - {1 \over 2}$. Then Theorem~\ref{thm-VII-2}, with $H_{ s} =H_{\mu s} $,
implies that
\begin{equation*} B_\theta (H_{\mu s} - \lambda - i 0)^{-1} B_\theta \in C^\nu
(\{ (\mu, \lambda) \in (I_0 \cap \mathbb{R}) \times E^{\mu}_{\delta_n}\}).
\end{equation*}
Since $D_{\delta_ n} \setminus D_{{\rho \over 8} \delta_n} \owns \mu
\rightarrow - H_{\mu u} \in E^{\mu}_{\delta_n},$ the latter equation
yields, in turn, that
$$B_\theta (H^{(0)}_\mu - i 0) B_\theta \in C^\nu (D_{\delta_n} \setminus D_{{\rho \over 2} \delta_n}),$$
which, due to Proposition VII.4, yields
\begin{equation} \label{VII.22}
B_\theta (H_g - \mu - i 0)^{-1} B_\theta \in C^\nu (D_{\delta_n}
\setminus D_{{\rho \over 2} \delta_n}).
\end{equation}

Let $\e_g$ be the solution to the equation $e(H_{\mu s}) = - H_{\mu
u}$ for $\mu$. By Theorem V.3, $0=e(H_{\e_g s})+H_{\e_g u}$ is the
ground state energy of the operator $H^{(0)}_{\e_g}$ and therefore,
by Theorem II.1, $\e_g$ is the ground state energy of the operator
$H_g$. In the lemma below we show that for $g$ sufficiently small
\begin{equation} \label{VII.23} D_{\delta_n} \setminus D_{{\rho \over 8} \delta_n},\
\forall n \ge 0,\ \mbox{cover}\ (\e_g, \e_g + {1 \over 18}
\rho_0),\end{equation} This together with \eqref{VII.22} implies the
statement of Theorem I.1. \qed
\begin{lemma}  \label{lemma-VII-4}For $g$ sufficiently small, \eqref{VII.23} holds.
\end{lemma}
\begin{proof}  We claim that for $g$ sufficiently small and for $n \ge
0$
\begin{equation} \label{VII.24} D (\e_g, {\rho_0 \over 4} \delta_n) \subset D_{\delta_n}.
\end{equation}
We prove this claim by induction in $n$.  We assume it is true for
$n \le j - 1$ and prove it for $n = j$. For $j = 0$, the induction
assumption is absent and so our proof of the induction step yields
also the first step.

We introduce the notation $e (\mu) : = e (H_{\mu s}).$  First we use
the relation $e(\e_g ) = - H_{\e_g u}$ to obtain
\begin{equation} \label{VII.25} | e (\mu) + H_{\mu u} | \le | e (\mu) -
e (\e_g) | + | H_{\e_g u} - H_{\mu u} |.
\end{equation}
Next, let $\Delta_0 E (\mu)$ be defined by the relation $H_{\mu u} =
: \rho^{-1}_0 (\mu - e_0) - \Delta_0 E (\mu).$  Then by
\eqref{eqn:35} and analyticity of $\Delta_0 E (\mu)$ in $I_0$, $|
\partial_{\mu} \Delta_0 E (\mu) | \le {\alpha_0/ \rho_0}$. The
last two relations imply
$$| H_{\e_g u} - H_{\mu u} |
= | \rho^{-1}_0 (\e_g - \mu) + \Delta_0 E (\e_g)$$
\begin{equation} \label{VII.26} - \Delta_0 E (\mu) |
\le \rho^{-1}_0 (1 + \alpha_0) | \e_g - \mu|.
\end{equation}

Recall the definition $E_{n}(\lambda, H_{\mu s}):=\big(
\cR_{\rho}^n(H_{\mu})\big)_u\equiv\la
\cR_{\rho}^n(H_{\mu})\ra_\Omega,\ \lambda =-H_{\mu u}$. Now we
estimate the first term on the r. h. s. of \eqref{VII.25}. Define
\begin{equation}
\Delta_n E(\lambda, H_{\mu s}):=E_{n}(\lambda, H_{\mu
s})-\rho^{-1}E_{n-1}(\lambda, H_{\mu s}). \label{eqn:53}
\end{equation}
It is shown in \cite{Sigal}, Eqns (V.24)-(V.25) that $e(H_s)$ satisfies the equation
\begin{equation}
e(H_s)= \sum_{i=1}^\infty\rho^i\Delta_i E(e(H_s), H_s),
\label{eqn:60}
\end{equation}
where the series on the right hand side converges absolutely by the
estimate
\begin{equation}
|\partial_{\lambda}^m \Delta_n E(\lambda)|\le\alpha_n (\frac{1}{12}
\rho^{n+1})^{-m}\ for\ n\le j\ \mbox{and}\ m= 0,1,
\label{eqn:54}
\end{equation}
shown in
\cite{FroehlichGriesemerSigal2009}. The relation \eqref{eqn:60} together with the
definitions $e (\mu) : = e (H_{\mu s})$ and $e (\e_g) : = e(H_{\e_g
s} ) = - H_{\e_g u}$ implies
\begin{equation}
e(\mu)= \sum_{i=1}^\infty\rho^i\Delta_i E(e(\mu), H_{\mu s})
\label{eqn:60a}
\end{equation} and
\begin{equation}
e(\e_g)= \sum_{i=1}^\infty\rho^i\Delta_i E(e(\e_g), H_{\e_g s}).
\label{eqn:60b}
\end{equation}

We estimate the difference between these series. 
%
It follows from the analyticity of $E_n (\lambda, H_{s})$ in $H_s$,
see \cite{FroehlichGriesemerSigal2009}, Proposition V.3, that
$\Delta_i E (\lambda, H_{\mu s})$ are analytic in $\mu \in
D_{\delta_i}, i \le j-1$. Now, by the induction assumption
$D_{\delta_i} \supset D(\e_g, {\rho_0 \over 4} \delta_i)$ for $i \le
j-1$. Hence using the Cauchy formula we conclude from \eqref{eqn:54}
that for $i \le j-1$
$$| \partial_\mu \Delta_i E (\lambda, H_{\mu s}) | \le
\frac{4 \alpha_i}{(1 - \rho) \rho_0 \delta_i}\ \mbox{on}\ D (\e_g,
{\rho_0 \over 4} \delta_{i }).$$ The latter estimate together with
\eqref{eqn:54} gives
$$\sum\limits_{i = 1}^\infty \rho^i | \Delta_i E (e (\mu), H_{\mu s})
- \Delta_i E (e (\e_g), H_{\e_g s}) |$$ $$ \le \sum\limits_{i =
1}^{j - 1} \rho^i ({\alpha_i \over \delta_i} | e (\mu) - e (\e_g) |
+ \frac{4 \alpha_i}{(1 - \rho) \rho_0 \delta_i}| \mu - \e_g |) + 2
\sum\limits_{i = j}^\infty \rho^i \alpha_i $$ $$\le 2 0 \alpha_1
 | e (\mu) - e (\e_g) |+  {80\alpha_1 \over  (1 -
\rho) \rho_0} | \mu - \e_g | + 4 \alpha_j \rho^j $$ on $D (\e_g,
{\rho_0 \over 4} \delta_j)$, where we used that $\delta_j = {1 \over
18} \rho^{j }$. This estimate together with the relations
\eqref{eqn:60a} and \eqref{eqn:60b} gives
\begin{equation} \label{VII.27}| e (\mu) - e (\e_g) | \le
{40\alpha_1 \over 1 - \rho} \delta_j + 160 \alpha_{j} \delta_j
\end{equation} in $D (\e_g, {\rho_0 \over 4} \delta_j)$, provided
$\alpha_1 \le {1 \over 40}$.  This estimate together with
\eqref{VII.25} and \eqref{VII.26} and the definition of
$D_{\delta_n}$ implies \eqref{VII.24} with $n=j$, provided
\begin{equation} \label{VII.28}{1 + \alpha_0 \over 4} +{4 0 \alpha_1
\over 1 - \rho} + 16 0 \alpha_0 \le 1.
\end{equation}
Remembering the definition of $\alpha_j$, we see that the latter
conditions can be easily arranged by taking $g$ sufficiently small.
This proves \eqref{VII.24}.

Next we show that for $g$ sufficiently small
\begin{equation} \label{VII.29}D_{\tau \delta_n} \subset D (\e_g, 1.5 \rho_0 \tau \delta_n),\
\mbox{where}\ \tau = O(1).\end{equation} The proof of this embedding
proceeds by induction in $n$ along the same lines as the proof of
\eqref{VII.24} given above. We have
$$| e (\mu) + H_{\mu u} | \ge | H_{\e_g u} - H_{\mu u} | - | e (\mu) - e (\e_g) |.$$
Again using the equality in \eqref{VII.26} and the estimates $|
\partial_\mu \Delta_0 E (\mu) | \le \rho^{-1}_0 \alpha_0$ and
\eqref{VII.27}, we find
$$| e (\mu) + H_{\mu u} | \ge \rho^{-1}_0 (1 -\alpha_0 -  {160\alpha_1 \over \rho (1 - \rho)})
 | \mu - \e_g | - 8 0 \rho^{-1}
\alpha_j  \delta_j$$ in $D_{\delta_j}$, provided $\alpha_1 \le {1
\over 40}$. Let $\mu \in D_{\tau \delta_j}$. Then $|e (\mu) + H_{\mu
u} | \le  \tau \delta_j$, which together with the previous estimate
gives
$$ | \mu - \e_g | \le \rho_0 (1 - \alpha_0 - {160 \delta_1 \over \rho (1 - \rho)})^{-1} (\tau + {80 \alpha_j
\over \rho}) \delta_j.$$ This yields \eqref{VII.29}, provided $g$ is
sufficiently small.

Embeddings \eqref{VII.24} and \eqref{VII.29} with $\tau = {\rho
\over 8}$ imply that
$$D_{\delta_n} \setminus D_{{\rho \delta_n \over 8 }} \supset D (\e_g, {\rho_0 \over 4}\delta_n)
 \setminus D (\e_g, {3\rho_0 \rho \over 16} \delta_n).$$
Since $\forall n, {\rho_0 \over 4} \delta_n >  {3\rho_0 \rho \over
16} \delta_{n - 1}$, the sets on the r. h. s. cover the interval
$(\e_g, \e_g + {1 \over 18} \rho_0)$ and therefore so do the sets on
the l.h.s. . Hence the lemma follows.
\end{proof}

\secct{Supplement: Background on the Fock space, etc}
\label{sect-SA}
%
%
Let $ \fh$ be either $ L^2 (\RR^3, \mathbb{C}, d^3 k)$ or  $ L^2
(\RR^3, \mathbb{C}^2, d^3 k)$. In the first case we consider $ \fh$
as the Hilbert space of one-particle states of a scalar Boson or a
phonon, and in the second case,  of a photon. The variable
$k\in\RR^3$ is the wave vector or momentum of the particle. (Recall
that throughout this paper, the velocity of light, $c$, and Planck's
constant, $\hbar$, are set equal to 1.) The Bosonic Fock space,
$\cF$, over $\fh$ is defined by
\begin{equation} \label{eq-I.10}
\cF \ := \ \bigoplus_{n=0}^{\infty} \cS_n \, \fh^{\otimes n} \comma
\end{equation}
where $\cS_n$ is the orthogonal projection onto the subspace of
totally symmetric $n$-particle wave functions contained in the
$n$-fold tensor product $\fh^{\otimes n}$ of $\fh$; and $\cS_0
\fh^{\otimes 0} := \CC $. The vector $\Om:=1
\bigoplus_{n=1}^{\infty}0$ is called the \emph{vacuum vector} in
$\cF$. Vectors $\Psi\in \cF$ can be identified with sequences
$(\psi_n)^{\infty}_{n=0}$ of $n$-particle wave functions,  which are
totally symmetric in their $n$ arguments, and $\psi_0\in\CC$. In the
first case these functions are of the form, $\psi_n(k_1, \ldots,
k_n)$, while in the second case, of the form $\psi_n(k_1, \lambda_1,
\ldots, k_n, \lambda_n)$, where $\lambda_j \in \{-1, 1\}$ are the
polarization variables.

In what follows we present some key definitions in the first case
only limiting ourselves to remarks at the end of this appendix on
how these definitions have to be modified for the second case. The
scalar product of two vectors $\Psi$ and $\Phi$ is given by
\begin{equation} \label{eq-I.11}
\la \Psi \, , \; \Phi \ra \ := \ \sum_{n=0}^{\infty}  \int
\prod^n_{j=1} d^3k_j \; \overline{\psi_n (k_1, \ldots, k_n)} \:
\vphi_n (k_1, \ldots, k_n) \period
\end{equation}

Given a one particle dispersion relation $\om(k)$, the energy of a
configuration of $n$ \emph{non-interacting} field particles with
wave vectors $k_1, \ldots,k_n$ is given by $\sum^{n}_{j=1}
\om(k_j)$. We define the \emph{free-field Hamiltonian}, $\hf$,
giving the field dynamics, by
%
\begin{equation} \label{eq-I.17a}
(\hf \Psi)_n(k_1,\ldots,k_n) \ = \ \Big( \sum_{j=1}^n \om(k_j) \Big)
\: \psi_n (k_1, \ldots, k_n) ,
\end{equation}
for $n\ge1$ and $(\hf \Psi)_n =0$ for $n=0$. Here
$\Psi=(\psi_n)_{n=0}^{\infty}$ (to be sure that the r.h.s. makes
sense we can assume that $\psi_n=0$, except for finitely many $n$,
for which $\psi_n(k_1,\ldots,k_n)$ decrease rapidly at infinity).
Clearly that the operator  $\hf$ has the single eigenvalue  $0$ with
the eigenvector $\Om$ and the rest of the spectrum absolutely
continuous.

With each function $\vphi \in \fh$ one associates an
\emph{annihilation operator} $a(\vphi)$ defined as follows. For
$\Psi=(\psi_n)^{\infty}_{n=0}\in \cF$ with the property that
$\psi_n=0$, for all but finitely many $n$, the vector $a(\vphi)
\Psi$ is defined  by
\begin{equation} \label{eq-I.12}
(a(\vphi) \Psi)_n (k_1, \ldots, k_n) \ := \ \sqrt{n+1 \,} \, \int
d^3 k \; \overline{\vphi(k)} \: \psi_{n+1}(k, k_1, \ldots, k_n).
\end{equation}
These equations define a closable operator $a(\vphi)$ whose closure
is also denoted by $a(\vphi)$. Eqn \eqref{eq-I.12} implies the
relation
\begin{equation} \label{eq-I.13}
a(\vphi) \Om \ = \ 0 \period
\end{equation}
The creation operator $a^*(\vphi)$ is defined to be the adjoint of
$a(\vphi)$ with respect to the scalar product defined in
Eq.~(\ref{eq-I.11}). Since $a(\vphi)$ is anti-linear, and
$a^*(\vphi)$ is linear in $\vphi$, we write formally
\begin{equation} \label{eq-I.14}
a(\vphi) \ = \ \int d^3 k \; \overline{\vphi(k)} \, a(k) \comma
\hspace{8mm} a^*(\vphi) \ = \ \int d^3 k \; \vphi(k) \, a^*(k)
\comma
\end{equation}
where $a(k)$ and $a^*(k)$ are unbounded, operator-valued
distributions. The latter are well-known to obey the \emph{canonical
commutation relations} (CCR):
\begin{equation} \label{eq-I.15}
\big[ a^{\#}(k) \, , \, a^{\#}(k') \big] \ = \ 0 \comma \hspace{8mm}
\big[ a(k) \, , \, a^*(k') \big] \ = \ \delta^3 (k-k') \comma
\end{equation}
where $a^{\#}= a$ or $a^*$.

Now, using this one can rewrite the quantum Hamiltonian $\hf$ in
terms of the creation and annihilation operators, $a$ and $a^*$, as
\begin{equation} \label{Hfa}
\hf \ = \ \int d^3 k \; a^*(k)\; \om(k) \; a(k) \comma
\end{equation}
acting on the Fock space $ \cF$.

More generally, for any operator, $t$, on the one-particle space $
\fh$ we define the operator $T$ on the Fock space $\cF$ by the
following formal expression $T: = \int a^*(k) t a(k) dk$, where the
operator $t$ acts on the $k-$variable ($T$ is the second
quantization of $t$). The precise meaning of the latter expression
can obtained by using a basis $\{\phi_j\}$ in the space $ \fh$ to
rewrite it as $T: = \sum_{j} \int a^*(\phi_j) a(t^* \phi_j) dk$.

To modify the above definitions to the case of photons, one replaces
the variable $k$ by the pair $(k, \lambda)$ and adds to the
integrals in $k$ also the sums over $\lambda$. In particular, the
creation and annihilation operators have now two variables: $a_
\lambda^\#(k)\equiv a^\#(k, \lambda)$; they satisfy the commutation
relations
\begin{equation} \label{eq-I.15}
\big[ a_{\lambda}^{\#}(k) \, , \, a_{\lambda'}^{\#}(k') \big] \ = \
0 \comma \hspace{8mm} \big[ a_{\lambda}(k) \, , \,
a_{\lambda'}^*(k') \big] \ = \ \delta_{\lambda, \lambda'} \delta^3
(k-k') .
\end{equation}
One can also introduce the operator-valued transverse vector fields
by
$$a^\#(k):= \sum_{\lambda \in \{-1, 1\}} e_{\lambda}(k) a_{\lambda}^\#(k),$$
where $e_{\lambda}(k) \equiv e(k, \lambda)$ are polarization
vectors, i.e. orthonormal vectors in $\mathbb{R}^3$ satisfying $k
\cdot e_{\lambda}(k) =0$. Then in order to reinterpret the
expressions in this paper for the vector (photon) - case one either
adds the variable $\lambda$ as was mentioned above or replaces, in
appropriate places, the usual product of scalar functions or scalar
functions and scalar operators by the dot product of
vector-functions or vector-functions and operator valued
vector-functions.



\vspace{3mm}   \noindent {\bf Acknowledgements:}

A part of this work was done while the third author was visiting
ETH Z\"urich, ESI Vienna and IAS Princeton. He is grateful to these institutions
for hospitality.
 \vspace{3mm}


\end{document}